\documentclass[11pt,twoside]{atmp}

\usepackage{latexsym}
\usepackage{epsfig,amssymb,euscript, mathrsfs}
\usepackage{amsmath}
\usepackage[all]{xy}
\usepackage{color}

\def\be{\begin{equation}}
\def\ee{\end{equation}}
\def\bea{\begin{eqnarray}}
\def\eea{\end{eqnarray}}

\def\cal{\mathcal}

\newcommand{\nn}{\nonumber}

\newcommand\R{\mathbb{R}}
\newcommand\Z{\mathbb{Z}}

\newcommand\C{\mathbb{C}}

\newcommand\diff{\mathrm{d}}
\newcommand{\de}{\partial}

\newcommand{\dd}{\mathrm{d}}
\newcommand{\me}{\mathrm{e}}
\newcommand{\ii}{\mathrm{i}}

\newcommand{\rot}{\mathcal{R}}

\begin{document}

\title[Non-K\"ahler heterotic rotations]
{Non-K\"ahler heterotic rotations}

%\arxurl{<hep_reference_#>}

\author[D. Martelli and J. Sparks]{Dario Martelli$^1$ and James Sparks$^2$}

\address{$^1$ Department of Mathematics, King's College London, \\
The Strand, London WC2R 2LS,  United Kingdom\\
\vskip 0.5cm
$^2$ Mathematical Institute, University of Oxford,\\
24-29 St Giles', Oxford OX1 3LB, United Kingdom}  %lines should be separated with double backslashes: \\
\addressemail{dario.martelli@kcl.ac.uk, sparks@maths.ox.ac.uk}

\begin{abstract}
We study a supersymmetry-preserving solution-generating method in heterotic supergravity. 
In particular, we use this method to construct one-parameter non-K\"ahler deformations of Calabi-Yau
manifolds with a $U(1)$ isometry, in which the complex structure remains invariant. 
We explain how to obtain corresponding solutions to heterotic string theory, up 
to first order in $\alpha'$, by a modified form of the standard embedding. 
In the course of the paper we also show that Abelian heterotic supergravity embeds into 
type II supergravity, and note that the solution-generating method in this context is related to a dipole-type 
deformation when there is a field theory dual.
\end{abstract}

\maketitle

%\cutpage %move this line so that the first page breaks at the appropriate place.

%\setcounter{page}{<insert page # for second page>}
%%%%%%%%%%%%%%%%%%%%%%%%%%%%%%%%%%%%%%%%%%%%%%%%%%%%%%%%%%%%%

\section{Introduction}

Supersymmetric solutions of the common sector of supergravity theories in ten dimensions have 
attracted the attention of both the  {\tt hep-th} and {\tt math.DG} communities. 
In particular, 
in six real dimensions these have become known as \emph{non-K\"ahler} geometries. 
They are one of the simplest examples of geometries characterized by a $G$-structure, 
which arise naturally in supergravity theories  \cite{Gauntlett:2002sc}. 
From the differential geometry point of view, the interest comes from the fact that the relevant 
$SU(3)$ structure is a relatively simple modification of $SU(3)$ holonomy, 
the latter characterizing Calabi-Yau three-folds. 
Essentially one replaces the Levi-Civita connection by a connection with skew-symmetic torsion. 
The manifolds are then still complex, but not symplectic, 
hence in particular they are not K\"ahler manifolds \cite{Strominger:1986uh,Hull:1986kz}. 
These structures are, however, quite rich;  for example, they are relevant for studies of mirror symmetry 
\cite{reidfantasy,mirrors} and  conifold transitions \cite{Maldacena:2009mw,transit}.  
Other work on heterotic non-K\"ahler geometries includes, for example, \cite{morehetero}.
This is also the appropriate framework for studying supersymmetric configurations of
 five-branes, and hence it features in the gauge/gravity duality.  
Most notably, a non-K\"ahler solution related to ${\mathcal N}=1$ SYM theory was discussed in \cite{Maldacena:2000yy}.
It is worth emphasizing that there are also generalizations of the non-K\"ahler equations to other dimensions. For example, in seven and eight dimensions the  geometries are characterized by torsion-full $G_2$ and $Spin(7)$ structures, 
respectively \cite{Gauntlett:2002sc,Gauntlett:2003cy,friedivanov,ugran}.
Most of our results apply also to these cases.

In the context of heterotic or type I string theory, there is also 
a non-Abelian  $SO(32)$ or $E_8\times E_8$ gauge field $\mathscr{A}$.
In the latter case, the non-Abelian gauge field plays a crucial role 
in constructions of string theory vacua that are designed to reproduce semi-realistic 
four-dimensional particle physics \cite{Witten:1985bz,Wittens,GSW2}. 
Vacua in which the internal field strength
has a background expectation value transforming in a subgroup $G$ of one of the  $E_8$ factors 
give as four-dimensional gauge group the commutant of $G$ in $E_8$. In the simplest 
construction, known as the standard embedding,  one identifies the internal gauge field 
with the $SU(3)$ spin connection of a background Calabi-Yau metric, giving $E_6$
as the observed gauge group. 
In the first part of this paper we will mainly focus on a $U(1)$ gauge field inside the common 
Abelian subgroup $U(1)^{16}$ of $SO(32)$ and $E_8\times E_8$; we shall then later discuss how to incorporate non-Abelian gauge fields 
that may lead to four-dimensional gauge groups $SO(10)$ and $SU(5)$, using a modified form of 
the standard embedding.

The mathematical understanding of non-K\"ahler geometries is much less developed than that of Calabi-Yau manifolds. 
Despite the very impressive existence results obtained recently in \cite{yau}, it is still desirable to construct new explicit examples. 
%This is particularly so for applications to gauge/gravity duality. 
The reformulation of the supersymmetry conditions in terms of equations for an $SU(3)$ structure 
 \cite{Strominger:1986uh,Hull:1986kz,Lopes Cardoso:2002hd,Gauntlett:2003cy}
is in fact not particularly helpful for this task, as one might have hoped. 
To make progress, the standard approach is that of making an \emph{ansatz} for the metric 
(or $G$-structure) enjoying additional symmetries. 
In this paper we will explore a different method for constructing new solutions, namely a 
\emph{solution-generating} transformation. 
Generally, these transformations exploit the symmetries of supergravity theories, and here we 
will discuss how such methods can be adapted to study solutions to the 
non-K\"ahler equations, and their cousins in dimensions $d\neq 6$.  

Buscher T-duality is at the heart of most of the generating techniques in supegravity. 
A simple transformation involving T-duality is the so-called TsT transformation. 
This has been applied to a variety of backgrounds, 
producing interesting gravity duals. The basic idea is that 
this transformation induces a $B$-field on the two-torus on which it is performed, hence giving
a non-commutative deformation  \cite{Seiberg:1999vs}.\footnote{Another generating method, that uses a chain of U-dualities, was proposed in
\cite{Maldacena:2009mw,Gaillard:2010gy} and may also be interpreted 
as a certain non-commutative deformation.}  
Examples include the gravity duals of non-commutative field theories  \cite{Maldacena:1999mh}, the beta-deformations
 \cite{Lunin:2005jy}, dipole deformations  \cite{Bergman:2001rw},  and 
 non-relativistic (Schr\"odinger) deformations \cite{nonrelat}. 
 Essentially the same transformation (see appendix 
 \ref{TrTappendix}) was used earlier to construct configurations of 
D-brane bound states  \cite{Breckenridge:1996tt}. For a concise review, see  \cite{Imeroni:2008cr}. 

The TsT transformation is part of the  $O(d,d)$ T-duality 
group\footnote{See \cite{CatalOzer:2005mr} for an explicit calculation of the $O(2,2)$ transformation.}
of type II supergravities \cite{Maharana:1992my,Bergshoeff:1995cg}. 
In the heterotic theory the generalized T-duality group 
is enlarged to $O(d+16,d)$, where 16 is the dimension of the Cartan subgroup of the gauge group 
\cite{Hassan:1991mq}. Transformations with elements of this group have been employed in the literature to construct (non-supersymmetric) charged
black holes, starting from uncharged black holes \cite{Chow:2008fe,Sen:1991zi,Sen:1992ua,Cvetic:1996dt}.

The solution-generating method that we will discuss in this paper, in the context of heterotic supergravity, involves
a particular subgroup $O(2,1)$ of the $O(d+16,d)$ group. 
However, instead of using the $O(d+16,d)$ transformation rules, we will derive the transformation from a rather 
different point  of view, which is analogous to the TsT transformation.  
Starting from a heterotic solution with a $U(1)$ symmetry, which includes Calabi-Yau manifolds 
with a $U(1)$ isometry as a special case, essentially 
the transformation that we will employ consists of a  rotation of the coordinates on an auxiliary
two-torus, followed by an ordinary Buscher T-duality (not involving the gauge fields), 
and then an opposite rotation. We will refer to this as an rTr transformation. 
A more detailed description will be given shortly.  In particular, we will 
show  that this transformation \emph{preserves supersymmetry} of the heterotic 
theory,\footnote{The reader might think that this is guaranteed by T-duality. However, we have been somewhat cavalier in describing the procedure here. 
The subtleties involved will become clear later in the paper.}  hence in particular it preserves the non-K\"ahler equations. 

In the process we will also show that 
this procedure may be phrased entirely in terms of a simple $O(d,d)$
transformation in type II. In fact, we will see that physically 
the solution-generating method, once embedded into type II, is essentially 
equivalent to a TsT transformation.  
This can be applied also to  backgrounds which are non-K\"ahler to begin with. 
For example, one can apply it to gravity duals 
of field theories \cite{Maldacena:2000yy}. In this case our formulas then give the supergravity 
 duals of certain dipole deformations, that we briefly discuss at the end of the paper.  

For applications to heterotic string theory, the expansion in powers of $\alpha'$ becomes important. In particular, the Bianchi identity 
and equations of motion receive corrections which involve the \emph{gravitational} Chern-Simons terms and the full \emph{non-Abelian} heterotic 
gauge field, already at first order. However, the transformation that we will discuss does not act naturally on these terms. We will nevertheless 
explain how to restore $\alpha'$ in our transformation, and combine this with a certain ``modified standard embedding'', 
in order to obtain full solutions to first order in $\alpha'$, including a non-trivial non-Abelian gauge field. 

The rest of this paper is organized as follows. In section \ref{heteroticsection} we summarize and review heterotic supergravity.
In section \ref{generating} we discuss our solution-generating method and present general formulas for the transformation. 
Section \ref{nonka} focuses 
on the non-K\"ahler deformation of Calabi-Yau geometries. In section \ref{alphaprime} we restore $\alpha'$ and incorporate 
the heterotic non-Abelian gauge field. In section \ref{theend} we discuss our results and possible directions for future work. 
Appendix \ref{integrability} contains a simple computation showing that compatibility of the supersymmetry equations and equations of motion imply that the curvature of the connection 
used in the modified Bianchi identity is an \emph{instanton}. In appendix \ref{TrTappendix} 
we write general formulas for TsT and TrT transformations (in type II), 
and note they are (locally) equivalent.

\section{The low-energy limit of heterotic strings}
\label{heteroticsection}

We begin by reviewing the low-energy limit of heterotic string theory. This is 
a heterotic supergravity theory with an $\alpha'$ expansion, where $1/2\pi\alpha'$ is the 
heterotic string tension. The reader might think that this is rather standard material. Although this should be the case, 
unfortunately there are many misunderstandings in the literature.

The low-energy limit of heterotic string theory is described by a ten-dimensional 
supergravity theory with metric $g$, dilaton $\Phi$, three-form $H$, and 
gauge field $\mathscr{A}$. For the $SO(32)$ heterotic theory 
we may take $\mathscr{A}$ to be in the fundamental representation; thus locally 
$\mathscr{A}$ is a one-form with values in purely imaginary, skew-symmetric 
$32\times 32$ matrices. 
There is no such representation for the $E_8\times E_8$ string, and 
in general one should replace the particular traces that follow by $1/30$ 
of the trace in the adjoint representation. Alternatively, 
this is equal to the trace in the fundamental representation 
when restricted to the $SO(16)\times SO(16)$ subgroup of 
$E_8\times E_8$ -- see, for example, \cite{GSW2}.

We begin by introducing the covariant derivatives $\nabla^\pm$ with torsion:
\bea
\nabla_i^\pm V^j &=& \nabla_i V^j \pm \tfrac{1}{2}H^j_{\ \, ik}V^k~,
\eea
where $V$ is any vector field and $\nabla$ denotes the Levi-Civita 
connection of $g$. These have totally skew-symmetric torsion $\pm H$, respectively. The string frame 
action, up  to 
two loops in sigma model perturbation theory, 
is\footnote{The convention we use for the $\alpha'$-dependent terms differs from some other conventions found in the literature; 
but different conventions involve only a trivial  redefinition of $\alpha'$. Note that,
 as usual in physics, the traces tr and generators $T^a$ of the various non-Abelian groups are defined so that
tr$(T^aT^b)$ is positive-definite.} \cite{hulltownsend}
\bea\label{hetstringaction}
S &=& \frac{1}{2\kappa_{10}^2} \int \dd^{10}x \sqrt{-g}\, \me^{-2\Phi} \Big[R   + 4(\nabla \Phi)^2 
- \tfrac{1}{12} H_{ijk}H^{ijk} \nn \\
&&\qquad \qquad \qquad\qquad\qquad - \alpha' \left( \mathrm{tr}\, \mathscr{F}_{ij} \mathscr{F}^{ij} - \mathrm{tr}\, \mathscr{R}^-_{\ ij} \mathscr{R}^{-\, ij}  \right) \Big] +
O(\alpha'^2)~.
\eea
The corresponding equations of motion are
\bea\label{hetstringEOM}
{R}_{ij}+2\nabla_i\nabla_j\Phi-\tfrac{1}{4}{H}_{ikl}{H}_{j}^{\ kl} - 2{\alpha'}\, \mathrm{tr}\, \mathscr{F}_{ik}\mathscr{F}_{j}^{\ k}\qquad \qquad &&\nn\\
 +2\alpha'\, {R}^-_{\ iklm}
{R}^{-\ \ klm}_{\ j} + O(\alpha'^2)&=&0~,\nn\\
\nabla^2(\me^{-2\Phi})-\tfrac{1}{6}\me^{-2\Phi}\, {H}_{ijk}{H}^{ijk} -{\alpha'}\, \me^{-2\Phi} \mathrm{tr}\,  \mathscr{F}_{ij}\mathscr{F}^{ij} \qquad &&\nn\\ 
+ {\alpha'}\, \me^{-2\Phi} \mathrm{tr}\,  \mathscr{R}^-_{\ ij}\mathscr{R}^{-\, ij}+O(\alpha'^2)&=&0~,\nn\\
\nabla^i\left(\me^{-2\Phi}\, {H}_{ijk}\right)+O(\alpha'^2)&=&0~,\nn\\
\nabla^{+\, i}\left(\me^{-2\Phi}\mathscr{F}_{ij}\right)+O(\alpha'^2)&=&0~.
\eea
Here $\mathscr{F}=\diff\mathscr{A}+\mathscr{A}\wedge\mathscr{A}$ is the field strength for $\mathscr{A}$, and ${R}^\pm_{\ ijkl}$ denotes the 
curvature tensor for $\nabla^\pm$. We shall denote the corresponding curvature two-forms 
by $\mathscr{R}^{\pm}_{\ ij}$, so that
\bea
\mathscr{R}_{\ ij}^{\pm \ ab} &=& R^{\pm}_{\ ijkl}e^{ak}e^{bl}~,
\eea
where $e^a_k$ is an orthonormal frame (vielbein).
The Bianchi identity for $H$ is 
\bea\label{hetstringBianchi}
\diff H &=& 2{\alpha'}\left(\mathrm{tr}\, \mathscr{F}\wedge\mathscr{F}-\mathrm{tr}\, \mathscr{R}^-\wedge \mathscr{R}^- \right)+O(\alpha'^2)~.
\eea
The corresponding supersymmetry equations are \cite{roo}
\bea\label{hetstringSUSY}
\nabla_i^+ \epsilon +O(\alpha'^2) & = & 0~,\nn \\
\left( \Gamma^i\de_i \Phi + \tfrac{1}{12} H_{ijk}\Gamma^{ijk} \right) \epsilon +O(\alpha'^2) & = & 0~, \nn\\
 \mathscr{F}_{ij} \Gamma^{ij} \epsilon +O(\alpha'^2) & = & 0~.
\eea
Here $\epsilon$ is a Majorana-Weyl spinor  and 
$\Gamma_i$ generate the Clifford algebra $\{\Gamma_i,\Gamma_j\}=2{g}_{ij}$. 
We shall refer to the last equation for $\mathscr{F}$ in (\ref{hetstringSUSY}) as  the \emph{instanton equation}.
That it is the curvature tensor for $\nabla^-$ that appears in the Bianchi identity (\ref{hetstringBianchi}), 
and then correspondingly the equations of motion (\ref{hetstringEOM}), was first noted by 
Hull \cite{Hull:1986kz}, and then subsequently discussed by other authors -- see, for example, \cite{Howe:1987nw}.
 This connection has \emph{opposite} sign torsion to the connection $\nabla^+$ 
appearing the gravitino equation in (\ref{hetstringSUSY}). Notice that one is free to make a field redefinition $H\leftrightarrow -H$, 
which then exchanges the roles of $\nabla^\pm$; in fact this opposite sign convention is also 
common in the literature. However, whatever the convention, there is always a \emph{relative} sign difference 
between the torsion of the connection
that appears in the gravitino equation and the connection that appears in the higher derivative terms. 
A common error in the literature is
to confuse these connections.
In the mathematics literature the connection 
$\nabla^+$ is  sometimes referred to as the \emph{Bismut connection}, 
discussed extensively in the context of the heterotic 
string, for example, in \cite{chern}. 
We shall refer to $\nabla^-$ as the \emph{Hull connection}.

Having briefly summarized the low-energy limit of heterotic string theory, in the next 
section we study a closely-related heterotic supergravity theory. Formally, this theory 
is obtained by (i) restricting to an Abelian subgroup of the gauge group, 
(ii) setting the higher order $O(\alpha'^2)$ corrections to zero in the  above 
theory, and then setting $\alpha'=1$, and (iii) formally setting to zero the higher derivative 
terms in $\mathscr{R}^-$. 
The solution-generating method we will present  in the next section
is for this Abelian theory, which we shall call \emph{Abelian heterotic supergravity}.\footnote{The 
Lagrangian and supersymmetry equations will be given explicitly in section \ref{generating}.} In 
particular, this theory does not contain $\alpha'$.
 We shall return to the heterotic string, and reinstate 
$\alpha'$, in section \ref{alphaprime}.

\section{Heterotic solution-generating method}
\label{generating}

In this section we present a supersymmetry-preserving solution-generating method in Abelian 
heterotic supergravity, 
where the initial solution has a $U(1)$ symmetry.
The construction relies on the observation that Abelian heterotic supergravity can 
be embedded supersymmetrically into type II supergravity. An $\mathrm{r}_2 \mathrm{T}\, \mathrm{r_1}$ transformation in type II theory, 
with an appropriate relation between the two rotations $\mathrm{r}_1$, $\mathrm{r}_2$, is then effectively a solution-generating 
transformation  in heterotic supergravity. 
We then show that this transformation lies in the $O(2,1)$ T-duality 
group of the heterotic theory itself. In fact the rotation is a close cousin 
of the Sen transformation, that takes the Kerr black hole to the Kerr-Sen black hole \cite{Sen:1992ua}. 
The latter was recently studied in higher dimensions in \cite{Chow:2008fe}. 

\subsection{Type II and heterotic supergravities}\label{procedure}

In this section we begin by showing that Abelian heterotic supergravity may be embedded supersymmetrically 
into the NS-NS sector of type II supergravity. This expands upon some comments made originally in 
\cite{Gauntlett:2003cy}. 

The NS-NS sector of type II supergravity is a ten-dimensional theory with 
metric $\hat{g}$, dilaton $\hat{\Phi}$, closed three-form $\hat{H}$, and string frame Lagrangian
\bea
\hat{\mathcal{L}} &=& \me^{-2\hat{\Phi}} \left(\hat{R}\, \hat{\star}\, 1 + 4\, \hat{\star}\, \diff\hat{\Phi}\wedge\diff\hat{\Phi} - \tfrac{1}{2}\, \hat{\star}\, \hat{H}\wedge \hat{H} \right)~.
\eea
The corresponding equations of motion are
\bea\label{TypeIIEOM}
\hat{R}_{ij}+2\hat\nabla_i\hat\nabla_j\hat\Phi-\tfrac{1}{4}\hat{H}_{ikl}\hat{H}_{j}^{\ kl}&=&0~,\nn\\
\hat\nabla^2(\me^{-2\hat\Phi})-\tfrac{1}{6}\me^{-2\hat\Phi}\, \hat{H}_{ijk}\hat{H}^{ijk}&=&0~,\nn\\
\hat\nabla^i\left(\me^{-2\hat\Phi}\, \hat{H}_{ijk}\right)&=&0~,
\eea
where $\hat{\nabla}$ denotes 
the Levi-Civita connection for $\hat{g}$.
The relevant supersymmetry equations are the gravitino and dilatino equations
\bea\label{TypeIISUSY}
\left(\hat{\nabla}_i  \pm \tfrac{1}{8}\hat{H}_{ijk}\hat{\Gamma}^{jk} \right) \epsilon_\pm & = & 0~, \nn\\
\left( \hat{\Gamma}^i\de_i \hat\Phi \pm \tfrac{1}{12} \hat{H}_{ijk}\hat{\Gamma}^{ijk} \right) \epsilon_\pm & = & 0~.
\eea
Here $\epsilon_\pm$ are Majorana-Weyl spinors  and 
$\hat{\Gamma}_i$ generate the Clifford algebra $\{\hat{\Gamma}_i,\hat{\Gamma}_j\}=2\hat{g}_{ij}$. 
A type IIA or type IIB solution will be supersymmetric if and only if there 
is at least one $\epsilon_+$ or $\epsilon_-$ satisfying (\ref{TypeIISUSY}), where 
$\epsilon_{\pm}$ have the opposite or same chirality, respectively.  
For the application we have in mind here we are interested in solutions 
with a single spinor $\epsilon_+$, which without essential loss of generality we take 
to have positive chirality. 

Consider such a supersymmetric solution which is invariant under a 
 Killing vector field $\partial_z$ which is nowhere zero. 
In particular, this should preserve the spinor $\epsilon_+$ 
under the spinor Lie derivative, as well as $\hat{\Phi}$ and $\hat{H}$. We may in general then write
\bea
\hat{g} &=& \me^{2\varphi}(\diff z+A_2)^2 + g~,\nn \\
\hat{\Phi} & = & \Phi + \tfrac{1}{2}\varphi~,\nn \\
\hat{H} &=& H + F_1\wedge (\diff z+A_2)~.
\eea
Here we have nine-dimensional fields, transverse to and invariant under $\partial_z$, 
comprising a metric $g$, scalar fields $\Phi$, $\varphi$, Abelian gauge fields $A_1$, $A_2$ with curvatures
$F_1=\diff A_1$, $F_2=\diff A_2$, and a three-form $H$. These satisfy 
equations of motion derived from the nine-dimensional Lagrangian
\bea\label{9dLag}
\mathcal{L} &=& \me^{-2\Phi}\Big(R\star 1  + 4\star\diff{\Phi}\wedge\diff{\Phi}- \star\,  \diff\varphi\wedge\diff\varphi \nn\\
&&- \tfrac{1}{2} \me^{-2\varphi} \star F_1 \wedge F_1 
-\tfrac{1}{2} \me^{2\varphi} \star F_2 \wedge F_2 - \tfrac{1}{2} \star {H}\wedge {H} \Big)~,
\eea
while the original Bianchi identity $\diff\hat{H}=0$ becomes 
$\diff H = -F_1\wedge F_2$. 
The supersymmetry equations (\ref{TypeIISUSY}) correspondingly reduce 
to equations for a nine-dimensional spinor $\psi$ satisfying 
\bea\label{9dSUSY}
\left(\nabla_\alpha + \tfrac{1}{8}H_{\alpha\beta\gamma}\sigma^{\beta\gamma}\right)\psi &=& \tfrac{\ii}{4}\left(\me^{-\varphi}F_{1\, \alpha\beta} + 
\me^{\varphi}F_{2\, \alpha\beta}\right)\sigma^\beta\psi~,\nn \\
\left(\sigma^\alpha\partial_\alpha \Phi + \tfrac{1}{12}H_{\alpha\beta\gamma}\sigma^{\alpha\beta\gamma}\right)\psi &=& 
\tfrac{\ii}{8}\left(\me^{-\varphi}F_{1\, \alpha\beta} + 
\me^{\varphi}F_{2\, \alpha\beta}\right)\sigma^{\alpha\beta}\psi~,\nn \\
\partial_\alpha\varphi\, \sigma^\alpha\psi &=& \tfrac{\ii}{4}\left(\me^{-\varphi}F_{1\, \alpha\beta} -
\me^{\varphi}F_{2\, \alpha\beta}\right)\sigma^{\alpha\beta}\psi~.
\eea
In deriving these equations we have used the fact that the ten-dimensional Killing spinor is invariant, ${\cal L}_{\de/\de z} \epsilon =0$. 
Here $\sigma_\alpha$ generate the Clifford algebra $\{\sigma_\alpha,\sigma_\beta\}=2g_{\alpha\beta}$, 
and we note that it is possible to choose conventions such that the Majorana condition in ten dimensions 
reduces to $\psi=\psi^*$, with the $\sigma_\alpha$ being symmetric and purely imaginary. 
We have taken a judicious linear combination of the reduced dilatino equation 
and $z$-component of the gravitino equation in presenting (\ref{9dSUSY}).

A T-duality in ten dimensions along 
$\partial_z$ is equivalent to a $\Z_2$ transformation of the 
nine-dimensional theory in which $\varphi\leftrightarrow -\varphi$, 
$A_1\leftrightarrow A_2$, with the other fields being invariant. 
In particular, notice that the supersymmetry equations (\ref{9dSUSY}) 
are manifestly invariant, with the last equation changing sign. This 
is a simple, direct proof that T-duality along such a direction 
preserves supersymmetry.

We also notice that a solution with $\varphi=0$ and $A_2=\mathcal{A}=-A_1$ 
has equations of motion that may be derived from the Lagrangian
\bea\label{9dhet}
\mathcal{L}_{\mathrm{Het}} &=& \me^{-2\Phi}\Big(R\star 1  + 4\star\diff{\Phi}\wedge\diff{\Phi}
-  \star \mathcal{F} \wedge \mathcal{F} - \tfrac{1}{2} \star {H}\wedge {H} \Big)~,
\eea
namely
\bea\label{hetEOM}
{R}_{\alpha\beta}+2\nabla_\alpha\nabla_\beta\Phi-\tfrac{1}{4}{H}_{\alpha\gamma\delta}{H}_{\beta}^{\ \gamma\delta} - \mathcal{F}_{\alpha\gamma}\mathcal{F}_{\beta}^{\ \gamma}&=&0~,\nn\\
\nabla^2(\me^{-2\Phi})-\tfrac{1}{6}\me^{-2\Phi}\, {H}_{\alpha\beta\gamma}{H}^{\alpha\beta\gamma} -\tfrac{1}{2}\me^{-2\Phi} \mathcal{F}_{\alpha\beta}\mathcal{F}^{\alpha\beta}&=&0~,\nn\\
\nabla^\alpha\left(\me^{-2\Phi}\, {H}_{\alpha\beta\gamma}\right)&=&0~,\nn\\
\nabla^\alpha\left(\me^{-2\Phi}\mathcal{F}_{\alpha\beta}\right)-\tfrac{1}{2}\me^{-2\Phi}\mathcal{F}^{\gamma\delta}H_{\beta\gamma\delta}&=&0~.
\eea
Here $\mathcal{F}=\diff \mathcal{A}$, the Bianchi identity is $\diff H = \mathcal{F}\wedge \mathcal{F}$, and the supersymmetry equations are
\bea\label{9dhetSUSY}
\left(\nabla_\alpha  + \tfrac{1}{8}H_{\alpha\beta\gamma}\sigma^{\beta\gamma} \right) \psi & = & 0~,\nn \\
\left( \sigma^\alpha\de_\alpha \Phi + \tfrac{1}{12} H_{\alpha\beta\gamma}\sigma^{\alpha\beta\gamma} \right) \psi & = & 0~, \nn\\
\mathcal{F}_{\alpha\beta} \sigma^{\alpha\beta} \psi & = & 0~.
\eea
These are the supersymmetry equations for 
a nine-dimensional heterotic supergravity theory, with 
Abelian gauge field $\mathcal{A}$ and curvature $\mathcal{F}=\diff\mathcal{A}$. More precisely, of course \emph{Abelian heterotic supergravity} 
exists in ten dimensions. This has the same Lagrangian, equations of motion and supersymmetry equations 
as (\ref{9dhet}), (\ref{hetEOM}) and (\ref{9dhetSUSY}), respectively, while in the present 
context the latter arise as a 
\emph{trivial} reduction of the ten-dimensional theory
to nine dimensions; that is, 
the ten-dimensional solution is assumed to be invariant 
under a \emph{constant length} Killing vector field, 
with all fields being transverse to (no ``legs'') and invariant under this direction. 
In particular, any supersymmetric solution to the above nine-dimesional theory 
trivially lifts to a supersymmetric ten-dimesional Abelian heterotic solution. 

We also see that any supersymmetric nine-dimesional heterotic solution 
lifts to a supersymmetric type II solution, via the ansatz
\bea\label{TypeIIembed}
\hat{g} &=& (\diff z+\mathcal{A})^2 + g~,\nn \\
\hat{\Phi} & = & \Phi~,\nn \\
\hat{H} &=& H - \mathcal{F}\wedge (\diff z+\mathcal{A})~.
\eea
It is in this sense that (nine-dimensional) Abelian heterotic supergravity is embedded 
into type II supergravity.

\subsection{The solution-generating rotation}
\label{generalrotate}

The idea in this section is to combine the observation of the previous section 
with certain standard solution-generating methods in type II. 
Since Abelian heterotic supergravity embeds into type II supergravity, 
it follows that the corresponding duality symmetries will also be related. 

We begin with a supersymmetric Abelian heterotic solution 
embedded into type II, via (\ref{TypeIIembed}). In addition we assume that the heterotic solution 
has a $U(1)$ symmetry that preserves the Killing spinor $\psi$. 
We may thus write the heterotic solution as
\bea\label{initialhet}
g &=& \frac{1}{V}(\diff\tau+A)^2 + Vg_\perp~,\nn\\
\mathcal{A} & = & a\, \diff\tau + \mathcal{A}_\perp~,\nn\\
B&=& B_1\wedge \diff\tau + B_\perp~,
\eea
where we have introduced the $B$-field via
\bea
H &=& \diff B + \mathcal{A}\wedge\diff\mathcal{A}~.
\eea
Here $\partial_\tau$ is the Killing vector field that generates the $U(1)$ 
symmetry. $V$ and $a$ are functions, 
$\mathcal{A}_\perp$, $B_1$ are one-forms, and 
$B_\perp$ is a two-form, all of which are invariant under $\partial_\tau$ 
and transverse to it.\footnote{That is, they are \emph{basic} forms with respect 
to $\partial_\tau$.} Notice that 
although $\mathcal{A}$ transforms under Abelian gauge transformations, 
$\tau$-independent gauge transformations leave $a$ gauge invariant. 

The equations in (\ref{TypeIIembed}) embed such a solution into type II 
supergravity. The corresponding type II solution now has \emph{two} 
symmetries, generated by $\partial_z$ and $\partial_\tau$. 
We may thus apply an $\mathrm{r}_2\mathrm{T}\, \mathrm{r}_1$ transformation. More precisely, 
we first rotate $(z,\tau)$ by an $SO(2)\cong U(1)$ rotation with constant angle $\delta_1\in [0,2\pi)$, 
then perform a T-duality along the new first coordinate $\tilde{z}$, and finally perform 
another rotation with constant angle $\delta_2\in [0,2\pi)$. These operations 
manifestly preserve supersymmetry in the type II theory. Having done this, 
we may then reduce back to nine dimensions on the final $z$ circle, to 
obtain a supersymmetric nine-dimensional solution to the theory with Lagrangian (\ref{9dLag}). We denote these $\mathrm{r}_2\mathrm{T}\, \mathrm{r}_1$-transformed fields 
with a prime.\footnote{We note that essentially this procedure was applied to 
higher-dimensional black hole solutions in \cite{Chow:2008fe}, with the rotations replaced 
by boosts so that $\partial_\tau$ is a timelike direction.}

In general such transformations do not preserve the original 
embedding of the heterotic theory into type II. However, 
an explicit calculation shows that the embedding is indeed 
preserved provided one takes $\delta_1=-\delta_2=\delta$. 
In particular, with this choice the 
final nine-dimensional rotated gauge fields ${A}'_1$, ${A}'_2$ for the theory with Lagrangian (\ref{9dLag}) obey 
${A}'_2=-{A}'_1$, while ${\varphi}'=0$. We may  denote this 
transformation more precisely as $\mathrm{r}^{-1}\mathrm{T}\, \mathrm{r}$, 
and the above procedure then leads to a supersymmetric heterotic solution given explicitly by
\bea\label{rotation}
{g}' &=& \frac{1}{V'}(\diff{\tau}'+{A}')^2 + V g_\perp~,\nn\\
\me^{2{\Phi}'} &= &  \frac{1}{h}\, \me^{2\Phi}~,\nn\\
{\mathcal{A}}'&=& {a}'\, \diff{\tau}' +\mathcal{A}'_\perp~,\nn\\
{B}'&=& {B}'_1\wedge \diff{\tau}' + {B}'_\perp~,
\eea
where we have defined the functions
\bea\label{rotateV}
h &\equiv & (c-sa)^2 + \frac{s^2}{V} ~,\nn\\
{V}' &\equiv & V h^2 \, = \, V \left[(c-sa)^2 + \frac{s^2}{V}\right]^2~,
\eea
and $c=\cos\delta$, $s=\sin\delta$.
The rotated fields are 
\bea\label{rotatea}
{a}' &=& -\frac{1}{h}\left[(c^2-s^2)a + cs \left(1-a^2 - \frac{1}{V}\right)\right]~,\nn \\
A' &=& (c-sa)^2 A + s(c-sa)\mathcal{A}_\perp + s(c\mathcal{A}_\perp + sB_1)~,\nn \\
\mathcal{A}'_\perp &=& -\frac{1}{h}\left[(c-sa)(c\mathcal{A}_\perp +sB_1) - \frac{s}{V} (s\mathcal{A}_\perp + (c-sa)A)\right]~,\nn \\
{B}'_1 &=& \frac{1}{h}\left[(c-sa)(cB_1-s\mathcal{A}_\perp) - \frac{s}{V} ( c\mathcal{A}_\perp - (s+ ca)A ) \right]~,\nn  \\
{B}'_\perp &=& B_\perp + \frac{s}{h}\left[(c-sa)B_1\wedge \mathcal{A}_\perp - \frac{1}{V}(c\mathcal{A}_\perp + sB_1)\wedge A\right]~.
\eea

It is not surprising that this procedure can be understood 
entirely within the original heterotic theory itself. Our initial heterotic solution (\ref{initialhet}) has 
a $U(1)$ symmetry generated by $\partial_\tau$ and a single Abelian gauge field $\mathcal{A}$, and in fact
 the above rotation is then embedded into the $O(2,1)$ duality group of this theory. 
Similar transformations were first investigated by Hassan and Sen 
\cite{Hassan:1991mq}, \cite{Sen:1992ua}. 
The above procedure effectively ``geometrizes'' this $O(2,1)$ transformation 
as an $O(2,2)$ transformation of type II, now with two symmetries
generated by $\partial_z$, $\partial_\tau$. As well as 
giving a first principles proof that the transformation preserves 
supersymmetry, the above construction also embeds it into type II theory, which we shall elaborate 
on later in section \ref{theend}. We note that 
in reference \cite{Hassan:1994mq} it was 
shown quite generally that such duality transformations preserve \emph{worldsheet} supersymmetry.

Let us see explicitly that (\ref{rotation}) indeed lies in the $O(2,1)$ duality group of the heterotic theory, following
\cite{Maharana:1992my} and \cite{Bergshoeff:1995cg}. 
We begin by making the change of variables
\bea\label{changeoneforms}
\mathcal{V}&=&\sqrt{2}\left(\mathcal{A}_\perp - aA\right)~,\nn\\
C &=& -B_1+ a\mathcal{A}_\perp - a^2 A~.\eea
We then form a triplet of Abelian gauge fields, defining
\bea
\mathbf{A}&=&
\left(\begin{array}{c} A \\ C \\ \mathcal{V}\end{array}\right)~.
\eea
It is then straightforward to check that the rotation (\ref{rotation}) acts on $\mathbf{A}$ as
the $O(2,1)$ matrix
\bea
\rot &=& \left(
\begin{array}{ccc}
 \cos^2 \delta  & -\sin^2 \delta  & \tfrac{1}{\sqrt{2}}  \sin 2 \delta \\
 -\sin^2 \delta & \cos^2 \delta  &  \tfrac{1}{\sqrt{2}}  \sin 2 \delta \\
 \tfrac{1}{\sqrt{2}}  \sin 2 \delta &  \tfrac{1}{\sqrt{2}}  \sin 2 \delta & -\cos 2 \delta
\end{array}
\right)~.
\eea
Here the metric $\eta$ preserved by $\rot$ is 
\bea
\eta &=&\left(\begin{array}{ccc} 0 & 1 & 0 \\ 1 & 0 & 0 \\
0 &0& 1\end{array}\right)~,\eea
so that $\rot^T \eta\, \rot = \eta$.
Notice that $\rot$ is symmetric, $\rot^T=\rot$, and also that $\rot^2 = 1$.
This latter  fact 
is consistent with the original construction of the duality transformation in type II, where $\rot = \mathrm{r}^{-1} \mathrm{T}_{\mathrm{Type \; II}} \, \mathrm{r}$.  We also note that $\det \rot=-1$, and hence the transformation is \emph{not} continuously connected to the identity, in general.\footnote{In the special case in which one begins with a Calabi-Yau solution with zero gauge field, it \emph{is} connected to the identity. We discuss this further in the next section.} 
In particular, notice that
\bea
\rot_{\delta=0}& =& \left(\begin{array}{ccc} 1 & 0 & 0 \\ 0 & 1 & 0 \\
0 &0& -1\end{array}\right)~,\nn \\
\rot_{\delta=\pi/2}& = & \left(\begin{array}{ccc} 0 & -1 & 0 \\ -1 & 0 & 0 \\
0 &0& 1\end{array}\right) \ =\  \mathrm{T}_{\mathrm{Het}} ~.
\eea
Thus for $\delta=0$ the transformed solution differs from the original solution by reversing the sign of 
the Abelian gauge field $\mathcal{F}$. Note this is an obvious discrete symmetry of heterotic supergravity.
We have also noted that $\rot_{\delta=\pi/2}$ is precisely \emph{heterotic T-duality}. In terms of the original 
variables, this is
\bea
{g}' &=& \frac{1}{V(a^2+\frac{1}{V})^2} \left[\diff\tau' + B_1 - a(\mathcal{A}_\perp-aA)\right]^2 + Vg_\perp~,\nn\\
\me^{2{\Phi}'} &=& \frac{1}{(a^2+\frac{1}{V})}\, \me^{2\Phi}~,\nn\\
\mathcal{A}' &=& \frac{1}{(a^2+\frac{1}{V})}\left[a\, (\diff\tau' + B_1) + \frac{1}{V}(\mathcal{A}_\perp - aA)\right]~,\nn\\
{B}' &=& B_\perp-\frac{1}{(a^2+\frac{1}{V})}(\diff\tau'+B_1)\wedge \left(a\mathcal{A}_\perp + \frac{1}{V}A\right) ~.
\eea
These are the heterotic Buscher T-duality rules of \cite{Bergshoeff:1995cg}. 

Next we turn to the transformation of the scalars. Again following \cite{Maharana:1992my} and \cite{Bergshoeff:1995cg}, we 
define a $3\times 3$ matrix $M$ via
\bea
M^{-1} &= &
\left(
\begin{array}{ccc}
\frac{1}{V}\left(1+a^2 V\right)^2 & -a^2 V & \sqrt{2} a \left(1+a^2 V\right) \\
 -a^2 V & V & -\sqrt{2} a V \\
 \sqrt{2} a \left(1+a^2 V\right) & -\sqrt{2} a V & 1+2 a^2 V
\end{array}
\right)~.
\eea
This is also symmetric and lies in $SO(2,1)$, namely $(M^{-1})^T \eta\, M^{-1}= \eta$ and $\det M=1$. It is straightforward to check that the rotation formulae (\ref{rotateV}), (\ref{rotatea}) are equivalent to
\bea
 (M')^{-1} &=& \rot\, M^{-1} \rot^T~.
 \eea
Finally, one can check that the two-form
\bea
\mathcal{B}&=& B_\perp - \tfrac{1}{2}B_1\wedge A + \tfrac{1}{2}a A\wedge \mathcal{A}_\perp
\eea
is invariant. 
One can write a reduced action, where one reduces along $\partial_\tau$, which is manifestly $O(2,1)$-invariant  \cite{Maharana:1992my}, \cite{Bergshoeff:1995cg}. Regarding the initial heterotic solution as ten-dimensional, in the above variables this takes the form
\bea
S & =&  \int \dd^9 x \sqrt{- g}\, \me^{-2\phi}\Big[R + 4(\dd \phi)^2 - \tfrac{1}{2} H^2   + \tfrac{1}{8} \mathrm{tr}\left(\de_\mu M^{-1} \de^\mu M \right) \nn\\&& - \tfrac{1}{2}{\bf F}^T M^{-1} {\bf F}\Big]
\eea 
where  $\phi=\Phi+\frac{1}{4}\log V$ and
\bea
H  &=& \dd \mathcal{B} +\tfrac{1}{2} {\bf A}^T\eta \wedge {\bf F}
\eea
is $O(2,1)$ invariant, and we have defined $\alpha^2\equiv \frac{1}{p!} \alpha_{i_1\cdots i_p}\alpha^{i_1\cdots i_p}$ for 
a $p$-form $\alpha$.

\subsection{The Calabi-Yau case}
\label{CYcase}

In this subsection we specialize to the case where the initial heterotic solution 
has $B=\mathcal{A}=0$ and $\Phi=\Phi_0$ is constant. In particular, this means that the initial metric $g$ has 
special holonomy, {\it e.g.} a product of Minkowski space with a Calabi-Yau manifold or a $G_2$ holonomy manifold. 

In this case the rotation (\ref{rotation}) simplifies to
\bea\label{rotatedCY}
{g}' &=& \frac{1}{V'}(\diff{\tau}'+\cos^2\delta A)^2 +{V}g_\perp~,\nn\\
\me^{2{\Phi}'} &= &  \frac{1}{h}\, \me^{2\Phi_0}~,\nn\\
\mathcal{A}'&=&-\frac{\sin\delta\cos\delta}{h} \left[\diff\tau' -\frac{1}{V}(\diff\tau'+A)\right]~,\nn\\
{B}'&=& -\frac{h\, \sin^2\delta}{V'} \diff\tau'\wedge A~,
\eea
where as before $V'=Vh^2$, but now $h$ simplifies to
\bea\label{hCYcase}
h& =& \cos^2\delta + \frac{\sin^2\delta}{V}~.
\eea

Recall that setting $\delta=\pi/2$ simply gives the T-dual heterotic solution. 
Provided $\delta\neq \pi/2$ we may introduce the new coordinate
$w=\tau'/\cos^2\delta$, and rewrite (\ref{rotatedCY}) slightly as
\bea\label{wcoords}
{g}' &=& \frac{\cos^4\delta}{V'}(\diff w + A)^2 + {V}g_\perp~,\nn\\
\me^{2{\Phi}'} &= &  \frac{1}{h}\, \me^{2\Phi_0}~,\nn\\
\mathcal{A}'&=&-\sin\delta\cos\delta\, \left[\diff w - \frac{h}{V'}(\diff w +A)\right]~,\nn\\
{B}'&=& -\frac{h\, \sin^2\delta\cos^2\delta}{V'} \diff w\wedge A~.
\eea
The curvatures are
\bea\label{curvatures}
{\mathcal{F}'} &=& \sin\delta\cos\delta\, \diff\left[\frac{h}{V'}(\diff w +A)\right]~,\nn\\
{H}' &=& \frac{h^2\, \sin^2\delta\cos^2\delta}{V'^2} (\diff w+A)\wedge F~.
\eea

Notice that a very special case of this construction is obtained by starting with flat ten-dimensional Minkowski spacetime.  
Even in this case our transformation produces a non-K\"ahler geometry with non-trivial $B$ and gauge field.  It would be interesting 
to study whether one can quantize strings in these backgrounds. 

\subsection{Global analysis}
\label{global}

The heterotic rotation we have described gives by construction a new set of fields which 
satisfy the supersymmetry equations, Bianchi identity and equations of motion. 
One can then ask whether the rotated solution is a \emph{regular}
supergravity solution. In the first part of this subsection we 
address this question for the simplified case in section \ref{CYcase}. 
We prove that for all $\delta\in [0,\pi/2)$ the rotated solution 
is a smooth supergravity solution, with the underlying manifold 
remaining the same. In this sense, the deformation is like the
beta deformation of Maldacena-Lunin \cite{Lunin:2005jy}, in that 
the parameter $\delta$ may be varied continuously to give 
a smoothly connected one-parameter family of supergravity solutions. 
We then discuss 
the more general rotation of section \ref{generalrotate}. As for the beta deformation of \cite{Lunin:2005jy} 
a general global anaylsis is more complicated, and we 
restrict ourselves here only to some brief comments on global issues 
in this more general setting.

We begin with (\ref{wcoords}), which is valid for $\delta\neq \pi/2$. 
Consider first the locus of points $\mathcal{M}_0\subset \mathcal{M}$ in the original spacetime $\mathcal{M}$ where $\partial_\tau$ 
is non-zero. Since we assumed that $\partial_\tau$ generates a $U(1)$ isometry, it 
follows that $\mathcal{M}_0$ will fibre over some orbifold 
$\mathcal{M}_0/U(1)$. The one-form $(\diff \tau+A)$ is then a connection one-form 
on the corresponding $U(1)$ principal (orbi-)bundle. Notice that 
on $\mathcal{M}_0$ the function $V$ is finite and strictly positive, and 
from (\ref{hCYcase}) the same is true of $h$, and thus also of
${V}'=Vh^2$. Since the one-form $(\diff w+A)$ 
also appears in the rotated metric in (\ref{wcoords}), it follows 
that the corresponding locus $\mathcal{M}'_0\subset \mathcal{M}'$, where $\partial_w$ is non-zero in the rotated spacetime $\mathcal{M}'$, 
is diffeomorphic to $\mathcal{M}_0$, and that the rotated metric is regular on $\mathcal{M}_0'\cong \mathcal{M}_0$ provided 
one takes the period of $w$ to be the \emph{same} as the period of $\tau$. The rotation 
then simply rescales the size of the  $U(1)$ fibre.

Let us now look at the locus of points where $\partial_\tau$ vanishes. 
Take some connected component $S\subset \mathcal{M}$. By definition, $V$ will diverge 
along $S$ -- the way in which it diverges is crucial 
for the regularity of the metric in a neighbourhood of $S$. 
Essentially, the metric restricted to the normal directions 
to $S$ should approach the flat space metric in some polar coordinate 
system. It turns out we will not need to enter into the precise
details of this metric regularity condition. To see this, notice that near to $S$ we have 
to leading order $h\sim \cos^2\delta$, and ${V}'\sim \cos^4\delta\, V$. 
Thus near to $S$ the rotated metric is to leading order
\bea
{g}' &\approx & \frac{1}{V}(\diff w+A)^2 + Vg_\perp~.
\eea
This is \emph{identical} to the initial metric, with $\tau$ replaced 
by the new coordinate $w$ (which forms the diffeomorphism of $\mathcal{M}_0$ onto $\mathcal{M}'_0$). Since these have the same period, and 
since the metrics are the same to leading order near to $S$, it follows 
that the rotated metric will be regular at $S$ provided the initial 
metric is. In fact an alternative way to argue this is to note quite generally that
\bea
g' &=& g - \frac{\sin^2 \delta}{V} \frac{(\sin^2 \delta + 2 V \cos^2 \delta)}{(\sin^2 \delta +  V \cos^2 \delta )^2} (\dd w + A)^2~.
\eea
Then where $V\to 0$ we have to leading order
\bea
g' & \approx & g  - \frac{\sin^2 \delta}{V} (\dd w + A)^2~.
\eea
The second term is a global one-form on $\mathcal{M}'$, as discussed in more detail below, and hence the rotated metric is regular at these loci.
 Also notice that since $\me^{2{\Phi}'}=\frac{1}{h}\, \me^{2\Phi_0}$, 
and $h$ is smooth and nowhere zero, the rotated dilaton 
is smooth. 

We conclude that for all $\delta\in[0,\pi/2)$ the underlying 
manifold is the same, $\mathcal{M}'\cong \mathcal{M}$, and the rotated metric is smooth provided 
the initial metric was. Notice that, in general, this 
is not true for the T-duality limit $\delta=\pi/2$. 
For example, a codimension four fixed point set $S$ of 
$\partial_\tau$ becomes a five-brane in the T-dual solution. 
In this limit $h=\frac{1}{V}$ and ${V}'=\frac{1}{V}$. In particular, 
along $S$ the dilaton $\me^{2{\Phi}'}=\frac{1}{h}\, \me^{2\Phi_0}=V\, \me^{2\Phi_0}$ 
diverges to infinity, as expected for a five-brane supergravity 
solution. Of course, in this case the T-dual solution has no 
gauge field: ${\mathcal{A}'}=0$. 

Let us now turn to the rotated curvatures in (\ref{curvatures}). 
First note that $(\diff w+A)$ is a global smooth one-form 
on $\mathcal{M}'_0\cong \mathcal{M}_0$. The one-form $\frac{h}{V'}(\diff w+A)$ 
is then a global smooth one-form on the whole spacetime 
$\mathcal{M}'\cong \mathcal{M}$, since $\frac{h}{V'}$ smoothly tends to zero precisely 
along the loci where $\partial_w$ vanishes. More precisely, 
$\frac{h}{V'}$ vanishes as $r^2$, where $r$ is the distance 
to $S$, while $(\diff w+A)$ approaches a global one-form on the normal sphere bundle
to $S$ in $M$. Thus ${\mathcal{F}'}$ is actually an \emph{exact}
two-form. There is hence no quantized flux associated to 
the rotated gauge field, as one might expect given that 
$\delta$ may be turned on continuously. Similar comments apply also 
to ${H}'$, showing that it is a smooth global three-form on $\mathcal{M}'$. 
Notice that in the T-dual limit at $\delta=\pi/2$, the 
change in the global structure of the spacetime 
due to a five-brane would then similarly be accompanied by 
a delta function source in the Bianchi identity for $H'$.

This completes our discussion of the global regularity of the 
solutions in section \ref{CYcase}. We conclude this section 
by commenting briefly on the more general rotations in 
(\ref{rotation}). Again, for $\delta\neq \pi/2$ 
one can replace $\tau' = w\cos^2\delta$, and 
similar analysis shows that the new connection 
one-form on $\mathcal{M}'_0$ is now (see (\ref{changeoneforms}))
\bea
\diff w + A + \sqrt{2}\tan\delta\,\mathcal{V}-\tan^2\delta \, C~.
\eea
The (orbi)-bundle $\mathcal{M}'_0\rightarrow \mathcal{M}'_0/U(1)$ will then have the same 
topology as $\mathcal{M}_0\rightarrow \mathcal{M}_0/U(1)$, for all values of $\delta\in[0,\pi/2)$, only if 
$\mathcal{V}=\sqrt{2}(\mathcal{A}_\perp - aA)$ and $C$ (equivalently $B_1$) are \emph{global} one-forms on $\mathcal{M}_0/U(1)$. 
{\it A priori}, this need not be the case. One can similarly analyse the metric 
near to $S$. In this case the rotated metric is to leading order
\bea\label{smoothly}
{g}' &\approx & \frac{1}{V}(\diff w + A + 2\tan\delta\mathcal{A}_\perp)^2 + Vg_\perp~.
\eea
Crucial here is that, as well as $1/V$ going to zero along $S$, 
one also has that $a$ and $B_1$ tend to zero along $S$. 
This is because near to $S$ the coordinate $\tau$ 
is an angular coordinate on the sphere linking 
$S$ in $\mathcal{M}$, so that $\diff \tau$ is in fact singular at $S$. 
It follows that the coefficient of $\diff\tau$ in any smooth 
form must go to zero along $S$, at an appropriate rate.  
The above formula (\ref{smoothly}) then guarantees that the rotated metric 
is smooth at $S$ if the initial metric is.

\section{Non-K\"ahler geometries}
\label{nonka}

As first derived in \cite{Strominger:1986uh,Hull:1986kz}, 
a supersymmetric heterotic 
supergravity solution of the Poincar\'e-invariant form $\R^{1,3}\times X$ 
implies that the six-manifold $X$ has a canonical $SU(3)$ structure, satisfying the system of equations
\bea\label{Omegaeqn}
\diff \left(\me^{-2\Phi}\Omega\right)&=&0~,\\\label{omegaeqn}
\diff\left(\me^{-2\Phi}\omega\wedge\omega\right)&=&0~,\\\label{Hflux}
\ii(\bar\partial-\partial)\omega&=&H~.
\eea
Here $\omega$ and  $\Omega$ are the $SU(3)$-invariant real two-form and complex three-form,
respectively. In particular, equation (\ref{Omegaeqn}) implies that 
the associated almost complex structure is integrable, so that 
$X$ is a complex manifold with zero first Chern class. The $\partial$ and $\bar\partial$ 
operators in (\ref{Hflux}) are then the usual Dolbeault operators.
On the other hand, in general $\omega$ is not closed and 
hence $X$ is not a K\"ahler manifold. Such structures 
are now commonly referred to as \emph{non-K\"ahler geometries}, 
even though this nomenclature is somewhat vague. We note 
that very similar results hold for products 
of Minkowski space with complex $n$-folds, and 
also for other $G$ structures,  such as 
$G_2$-structure manifolds \cite{Gauntlett:2001ur,Gauntlett:2002sc,Gauntlett:2003cy}. 
Our analysis in this section should extend appropriately 
to all of these cases, but we content ourselves here with the most 
interesting case of an $SU(3)$ structure.

For Abelian heterotic supergravity, with no $\alpha'$, 
the supersymmetry equations (\ref{Omegaeqn}), (\ref{omegaeqn}), (\ref{Hflux}) 
are also supplemented by the gauge field equations
\bea\label{F11}
\mathcal{F}\wedge \Omega&=&0~,\\\label{HYM}
\omega\, \lrcorner\, \mathcal{F}&=&0~.
\eea
The first equation (\ref{F11}) says that the $U(1)$ gauge field strength $\mathcal{F}$ has Hodge type $(1,1)$, while (\ref{HYM})
is the \emph{Hermitian-Yangs-Mills} (HYM) equation for the gauge field. These are equivalent to the instanton equation for 
$\mathcal{F}$.
The Bianchi identity is
\bea\label{Bianchi}
\diff H &=& \mathcal{F}\wedge\mathcal{F}~.
\eea
In particular, the latter then ensures that the equations of motion are satisfied (see also 
appendix \ref{integrability}). 

\subsection{The rotated non-K\"ahler equations}

Starting with a supersymmetric $\R^{1,3}\times X$ heterotic supergravity solution in ten dimensions, 
the $O(2,1)$ transformation of section \ref{generating} produces a new 
$\R^{1,3}\times {X}'$ solution. If $X$ is a Calabi-Yau three-fold,
then we have already shown in general that provided $\delta\in[0,\pi/2)$ then
${X}'$ is diffeomorphic to $X$. However, for $\delta\neq 0$  ${X}'$ is equipped with a 
new $SU(3)$ structure, which is non-K\"ahler. In this 
section we write explicit formulas for the rotated $SU(3)$ structure, 
and briefly demonstrate how the $SU(3)$ structure equations, HYM equation and Bianchi identity 
are satisfied. We will be particularly interested in the transformation of the 
holomorphic $(3,0)$-form $\Omega$ in the next subsection.

We now suppress the Minkowski space directions, which may be
 absorbed into the rotation-invariant $Vg_\perp$ and play no role, and denote the initial Calabi-Yau three-fold 
metric by
\bea
g &=& \frac{1}{V}(\diff\tau +A)^2 + Vg_\perp~,
\label{initcy}
\eea
where $g_\perp$ is now a five-dimensional transverse metric. The initial K\"ahler form and holomorphic
$(3,0)$-form are, in a hopefully obvious notation,
\bea
\omega &=& e_\perp \wedge (\diff \tau + A) + V\omega_\perp~,\nn\\\label{Omega}
\Omega &=& (V e_\perp + \ii (\diff \tau + A)) \wedge V^{1/2} \Omega_\perp~.
\eea
Since the initial Killing spinor is assumed to be invariant under $\partial_\tau$, 
it follows that so are $\omega$ and $\Omega$. 
Notice that $Ve_\perp = J(\diff\tau +A)$, where the complex structure, metric and 
K\"ahler form are related via $J^i_{\ j} = g^{ik}\omega_{kj}$. With this sign convention, 
$\diff^c f \equiv J\diff f=\ii(\partial-\bar\partial)f$ acting on functions $f$. This sign convention 
is opposite to much of the mathematics literature, but is more common in the physics literature.

The closure of $\omega$ and $\Omega$ immediately lead to 
\bea\label{operp}
\diff e_\perp = 0~, \qquad \diff (V^{1/2}\Omega_\perp)=0~.
\eea
Denoting rotated quantities with primes as before, we have
\bea\label{rotatedOmega}
\me^{-2\Phi'} \Omega' &=& \cos^2\delta\, \Omega + \sin^2\delta \, e_\perp \wedge V^{1/2}\Omega_\perp~, 
\eea
which we immediately see is closed, as required by (\ref{Omegaeqn}). In these equations 
one should understand that we have formally replaced $\tau$ by $w$ in unrotated quantities such as $\Omega$ 
-- recall this is the diffeomorphism between $X$ and ${X}'$. 
Using the explicit form 
of ${\mathcal{F}'}$ in (\ref{curvatures}) it is also straightforward to check that 
${\mathcal{F}'}\wedge\Omega'=0$, so that ${\mathcal{F}'}$ is type $(1,1)$. 

The rotated two-form is
\bea
\me^{-2\Phi'}\omega' &=& \cos^2\delta\, \omega + \sin^2\delta\, \omega_\perp~,
\eea
so that
\bea
\me^{-2\Phi'} \omega' \wedge \omega' & = & \cos^2 \delta \, \omega \wedge \omega + \sin^2 \delta\, V \omega_\perp \wedge \omega_\perp~.
\eea
Hence 
\bea\label{dJeqn}
\diff ( \me^{-2\Phi'} \omega' \wedge \omega' ) &=& 0 \qquad \Leftrightarrow \qquad \diff ( V \omega_\perp \wedge \omega_\perp ) =0~.
\eea
But this follows immediately using (\ref{operp}) and 
\bea
\omega_\perp \wedge \omega_\perp &=& \frac{1}{2} \Omega_\perp \wedge \bar \Omega_\perp~,
\eea
which in turns follows from $\tfrac{1}{3!}\omega^3 = \tfrac{\ii}{8}\, \Omega\wedge\bar\Omega$.

Next we compute
\bea\label{HYMtoshow}
\qquad \me^{-4\Phi'}  \omega' \wedge \omega' \wedge {\cal F}' & =&  \sin\delta\cos^3 \delta \, \omega_\perp \wedge (\diff w
+ A) \wedge \Big[\diff V \wedge \omega_\perp \nn \\&& - 2 e_\perp \wedge F \Big]~.
\eea
The right hand side of (\ref{HYMtoshow}) vanishes on using
\bea
\diff \omega = 0 \qquad \Leftrightarrow \qquad \diff (V \omega_\perp ) = e_\perp \wedge F~,
\eea
together with (\ref{dJeqn}), which thus establishes the HYM equation (\ref{HYM}).

Finally, we compute
\bea
\diff\omega' &=& \diff \left(\omega - {\sin^2\delta}\, V'^{-1} e_\perp \wedge (\diff w+A)\right)~,\nn\\
&=& \tan \delta\, e_\perp \wedge {\cal F}'~.
\eea
For a $(1,1)$-form $\omega$ we have $\diff^c \omega 
= J\circ \diff \omega$, since $J\circ \omega = \omega$. 
Using also the fact that ${\cal F}'$ is type $(1,1)$ with respect to the transformed complex structure, it follows that 
\bea
\diff^c  \omega' &=& {J}' \circ \diff \omega' = \tan\delta\, {J}'(e_\perp) \wedge {\cal F}'~,\nn\\
&=& -\frac{h^2\, \sin^2\delta\cos^2\delta}{V'^2}\, (\diff w+A)\wedge F~,\nn \\
&=& -{H}'~.
\eea
Finally, in our conventions $\diff^c {\omega}' = \ii (\partial - \bar\partial)\omega'$, so we obtain
\bea
{H}'&=&\ii (\bar\partial - \partial)\omega'~,
\eea
which is equation (\ref{Hflux}).

The above provides an alternative proof that the rotation preserves supersymmetry 
in this particular case. As already mentioned, very similar computations 
could be done starting with other special holonomy manifolds. However, the main 
reason for presenting the above results is that we shall investigate 
in more detail how the complex structure and other pure spinor 
(in the sense of \cite{Hitchin}) $\me^{\ii\, \omega}$  
transform in the next section.

\subsection{Invariance of the complex structure}
\label{complex}

In this section we prove that the rotated complex structure, starting with a Calabi-Yau three-fold, 
is in fact invariant. Equivalently, ${\Omega}'$ is proportional to 
the original $\Omega$, after an appropriate diffeomorphism. 
This claim is not at all obvious from the formula (\ref{rotatedOmega}) for 
$\Omega'$ in terms of $\Omega$. 
Thus the heterotic rotation may be regarded as fixing 
the underlying complex manifold, but 
rotating the solution from K\"ahler to non-K\"ahler. 

In an appropriate coordinate patch we may introduce 
complex coordinates $z^0,z^1,z^2$ and a K\"ahler potential $K$ so that the 
initial Calabi-Yau metric is
\bea
g &=& 4\, \partial_a\bar{\partial}_b K\, \diff z^a \diff\bar{z}^b~,
\eea
where $a,b\in\{0,1,2\}$. We may also choose $z^0=x+\ii\tau$, 
where $\partial_\tau$ generates the $U(1)$ isometry, and 
correspondingly $\partial_\tau K=0$. In terms of 
our earlier notation, it is then straightforward to compute
\bea\label{stuff}
V = \frac{1}{\partial^2_x K}~, \qquad A = \frac{\ii}{\partial^2_xK}\left(\bar\partial_i\, \partial_xK\, \diff \bar{z}^i - \partial_i\, \partial_x K\, \diff z^i\right)~,
\eea
where $i=1,2$. Notice here that, although both $z^a$ and $K$ are defined only locally in each 
coordinate patch, with K\"ahler transformations acting on $K$ between patches, nevertheless 
the quantity $\partial^2_xK$ is a globally defined function on $X$. This is clear, 
since it is the square length of $\partial_\tau$, which is a globally defined  vector field by assumption. We also note that
\bea
e_\perp = \diff(\partial_xK)~.
\eea
Since $e_\perp$ is a global one-form on $X$, it follows
that $\partial_x K$ in different K\"ahler coordinate patches differ by an additive constant. Notice we 
may also write
\bea\label{Veperp}
Ve_\perp = \diff x + \frac{1}{\partial^2_xK}\left(\partial_i\, \partial_x K\, \diff z^i + \bar{\partial}_i\, \partial_x K\, \diff \bar{z}^i\right)~.
\eea

We now turn to the rotated holomorphic $(3,0)$-form, which after rewriting (\ref{rotatedOmega}) is
\bea
\me^{-2\Phi'}  \Omega' &=& \left[(V+\tan^2\delta)e_\perp + \ii (\diff w+A)\right]\wedge V^{1/2}\Omega_\perp~.
\eea
This then differs from $\Omega$ in (\ref{Omega}) only in the $\tan^2\delta$ term. Using the expression for 
$A$ in (\ref{stuff}) together with (\ref{Veperp}) this is 
\bea
\me^{-2\Phi'}  \Omega' &=& \left(\diff x + \tan^2\delta\, e_\perp + \ii \diff w + \frac{2}{\partial^2_xK}\partial_i\, \partial_x K\, \diff z^i\right)\wedge 
V^{1/2}\Omega_\perp~,\nn\\
&=& \left(\diff x + \tan^2\delta\, e_\perp + \ii \diff w\right)\wedge V^{1/2}\Omega_\perp~.
\eea
Here in the second line we have used that the closed complex two-form $V^{1/2}\Omega_\perp$ is proportional to 
$\diff z^1\wedge \diff z^2$. We thus see that $\me^{-2\Phi'}\Omega'$ is equal to the original $\Omega$ 
provided we make the coordinate change
\bea\label{coordchange}
{x}' &=& x + \tan^2\delta\, \partial_x K~.
\eea
For, then 
\bea
\me^{-2\Phi'} \Omega' &=& \left(\diff {x}'+ \ii\diff w\right)\wedge V^{1/2}\Omega_\perp~,
\eea
which is the same as the formula for $\Omega$ but with $x$ replaced by ${x}'$ and 
$\tau$ replaced by $w$. Recall that the latter is already part of the diffeomorphism 
between $X$ and ${X}'$ discussed in the previous sections. Notice that
in (\ref{coordchange}) the second term $\partial_x K$ on the right hand side is 
a globally defined smooth function, up to constant shifts between coordinate patches. 
The latter are  simply trivial constant shifts of $x$, and thus (\ref{coordchange}) defines 
a global diffeomorphism of $X$ and ${X}'$ which takes $\Omega$ into $\me^{-2\Phi'}{\Omega}'$.\footnote{An 
alternative argument is to note that if $b_1(X)=0$ then $e_\perp=\diff f$ for some 
\emph{global} function $f$, and then ${x}'=x +\tan^2\delta\, f$.}

The coordinate transformation (\ref{coordchange}) takes a particularly 
interesting form if the original Calabi-Yau metric is \emph{toric}, {\it i.e.} 
$U(1)^3$ invariant. In this case there is a symplectic coordinate system 
in which
\bea
\omega &=& \diff y^a\wedge \diff\phi^a~,
\eea
where without loss of generality we may take $\phi^0=\tau$. The corresponding 
complex coordinates, with $z^a=x^a+\ii \phi^a$, are related to the symplectic 
coordinates via
\bea
y^a &=& \frac{\partial K}{\partial x^a}~,
\eea
where one can take $K=K(x^a)$ to be $U(1)^3$-invariant.
Thus (\ref{coordchange}) may be rewritten
\bea
\cos^2\delta\, {x}' &=& \cos^2\delta\, x + \sin^2\delta\, y~,
\eea
where $y=y^0$ is the symplectic coordinate paired with $\tau=\phi^0$. 
In this sense, the coordinate transformation mixes complex and symplectic coordinates.

In summary, the rotation in fact preserves the underlying complex manifold.\footnote{In fact 
this had to be the case for toric manifolds. Here it is a standard fact that there is a unique 
complex structure that is compatible with the  torus action.}
However, it certainly changes the K\"ahler structure to a non-K\"ahler structure. 
In particular, we note that 
\bea
\exp({\ii\, {\omega}'}) & = & \exp\left[-\frac{1}{(1+\cot^2\delta\, V)}e_\perp \wedge (\diff w+A)\right] \wedge \exp({\ii\, \omega})~,
\eea
as usual with $\tau$ understood to be replaced by $w$ in $\omega$ on the right hand side.
Thus the rotated pure spinor $\me^{\ii\, \omega'}$ is related to the original 
pure spinor $\me^{\ii\, \omega}$ via a simple multiplying form. 
Compare this with the rotation in \cite{Maldacena:2009mw}, where the same pure spinor instead picks up a \emph{phase} 
in the rotated solution. 

\subsection{Examples}
\label{examples}

In this section we present simple explicit examples of non-K\"ahler deformations of Calabi-Yau geometries.
Namely, we discuss the Gibbons-Hawking and resolved conifold metrics.

\subsubsection*{The Gibbons-Hawking metric}\label{GH}

The Gibbons-Hawking metric is a hyper-K\"ahler metric in four real dimensions with a 
tri-holomorphic Killing vector field, {\it i.e.} a Killing vector field 
$\partial_\tau$ that preserves the triplet of complex structures implied by 
an integrable $SU(2)$ structure in four dimensions. The metric takes the well-known form
\bea
g &=& \frac{1}{V}(\diff\tau +A)^2 + V g_{\mathbb{E}^3}~,
\eea
where $V$ is any harmonic function on Euclidean three-space $\mathbb{E}^3$, and 
$\diff A = - *_3 \diff V$. By taking 
\bea\label{points}
V&=&\sum_{i=1}^m \frac{1}{|\mathbf{x}-\mathbf{x}_i|}~,
\eea
with $\mathbf{x}_i\in\mathbb{E}^3$ distinct points, this is the family of
asymptotically locally Euclidean metrics on the resolution of the $A_{m-1}$ singularity
$\C^2/\Z_m$. 

Applying the rotation, we obtain the string frame solution 
\bea\label{rotatedGH}
{g}' &=& \frac{1}{Vh^2}(\diff \tau' + \cos^2\delta\, A)^2 + Vg_{\mathbb{E}^3} \nn \\ & = & \frac{1}{{V}'}(\diff\tau' + \cos^2\delta\, 
A)^2+ {V} g_{\mathbb{E}^3}~,\nn\\
\me^{-2\Phi'} & = & h \, = \, \cos^2\delta + \frac{\sin^2\delta}{V}~,\nn\\
*{H}'&=&-\diff \log h~.
\eea
These satisfy a four-dimensional version \cite{Gauntlett:2003cy} of the equations (\ref{Omegaeqn}), (\ref{omegaeqn}), 
(\ref{Hflux}), with $\Omega$ now being the holomorphic $(2,0)$-form and equation (\ref{omegaeqn}) 
replaced by $\diff\left(\me^{-2\Phi'}{\omega}'\right)=0$. In the obvious orthonormal frame with
\bea
{e}'_0 \, = \, \frac{1}{\sqrt{V} h} (\diff \tau' + \cos^2\delta\, A)~, \qquad {e}'_i \, = \, \sqrt{V}\, \diff x_i~, \quad i=1,2,3~.
\eea
the gauge field curvature is given by
\bea
{\mathcal{F}'} &=& \frac{\sin\delta\cos\delta}{V^2 h} \left(\partial_i V\,{e}'_0\wedge {e}'_i - \tfrac{1}{2!} \epsilon^{ijk}
\partial_k V\, {e}'_i\wedge {e}'_j\right)~.
\eea
From this expression it is straightforward to check that ${\mathcal{F}'}$ is both type $(1,1)$ and satisfies 
the HYM equation $\omega'\, \lrcorner\,{\mathcal{F}'}=0$.
One also checks that the Bianchi identity $\diff {H}' = {\mathcal{F}'}\wedge {\mathcal{F}'}$ holds. 
In fact, more precisely this holds for all $\delta\in[0,\pi/2)$. For the T-duality limit with 
$\delta=\pi/2$ we have $\mathcal{F}'=0$, but the Bianchi identity 
for the solution with harmonic function (\ref{points}) then has 
$m$ delta-function sources on the right hand side; {\it cf}. the comments in section \ref{global}. 
This is as expected, since the T-dual solution is $m$ five-branes in flat spacetime, smeared over a circle $S^1$, and positioned 
at the points $\mathbf{x}_i\in\R^3$, where $\R^3$ is transverse to all the five-branes and the $S^1$ over which 
the branes are smeared. 

Any such solution in four dimensions is necessarily \emph{conformally} hyper-K\"ahler \cite{Gauntlett:2003cy}. 
This is clear from the second form of the metric in (\ref{rotatedGH}), 
since the rotated $Vh=\cos^2\delta\, V + \sin^2\delta$ is also harmonic, and thus of Gibbons-Hawking type.

\subsubsection*{The resolved conifold}\label{conifold}

Next, as an explicit example in dimension six, we consider the rotation of the Ricci-flat metric on the resolved 
conifold \cite{Candelas:1989js}. A similar analysis can be done for the Ricci-flat  K\"ahler metrics presented in
 \cite{Martelli:2007pv}. The resolved conifold metric can be written in the following explicit form 
\bea
\dd s^2_{\mathrm{RC}} & =  &\frac{1}{\kappa (r)}\dd r^2 + \frac{r^2}{6}\left( \dd \theta_1^2 + \sin^2 \theta_1 \dd \phi^2_1 \right) + \Big(\alpha^2 + \frac{r^2}{6}\Big)
\left(\dd \theta_2^2 + \sin^2 \theta_2 \dd \phi^2_2 \right)  \nonumber\\
&+ &\frac{r^2}{9}\kappa (r) (\dd \psi + \cos \theta_1 \dd \phi_1 + \cos \theta_2 \dd \phi_2 )^2~,
\eea
where 
\bea
\kappa (r)& =& \frac{9 \alpha^2 + r^2 }{6\alpha^2 + r^2}~.
\eea
The (asymptotically Reeb) Killing vector $\de /\de \psi$ does not leave the holomorphic $(3,0)$-form $\Omega$ invariant, and hence we can apply our rotation only to an arbitrary linear combination
of $\de /\de \phi_1$ and $\de /\de \phi_2$. We then first change coordinates, making a transformation
\bea\label{robotsindisguise}
\phi_1 &=& a \varphi_1 + b \varphi_2~, \nonumber \\
\phi_2 &=& c \varphi_1 + d \varphi_2 ~.
\eea
We will momentarily perform the 
rotation along $\de/\de\varphi_1$, without loss of generality, which recall is required to generate a $U(1)$ action. As such, the transformation (\ref{robotsindisguise}) should be taken to lie in $SL(2,\Z)$, 
although for the local computations that follow this comment is not important. 
We next write the metric adapted to the Killing vector $\de /\de \varphi_1$ in the form (\ref{initcy}), and compute
\bea
V^{-1} &=& a^2 \frac{r^2}{6} \sin^2 \theta_1 + c^2 \Big(\alpha^2 + \frac{r^2}{6}\Big)\sin^2 \theta_2 \nn \\&&+ \frac{r^2}{9}\kappa (r) (a \cos \theta_1 + c \cos\theta_2)^2~.
\eea
Notice that provided $c\neq 0$, this vanishes at the north ($\theta_2=\pi$) and south ($\theta_2=0$) poles of the resolved two-sphere. Of course, the resolved conifold
 metric is nevertheless perfectly smooth at these points. We also have
\bea
&&A \ =\  V \Bigg[ \Big( ab\, \frac{r^2}{6} \sin^2 \theta_1  + cd \Big(\alpha^2 + \frac{r^2}{6}\Big) \sin^2 \theta_2 + \nonumber\\
&&\frac{r^2}{9}\kappa (r) (a \cos \theta_1 + c \cos\theta_2) (b \cos \theta_1 + d \cos\theta_2)\Big) \dd \varphi_2 +\nonumber \\
&&\frac{r^2}{9}\kappa (r) (a \cos \theta_1 + c \cos\theta_2) \dd \psi \Bigg]~.
\eea

Let us now compute the volume of the resolved two-sphere as a function of the deformation parameter $\delta$. The induced metric at $r=0$ reads
\bea
\dd s^2 |_{S^2} & =&  \alpha^2 (\dd \theta_2^2 + f(\theta_2)  \sin^2 \theta_2 \dd \phi_2^2)~,
\label{deftwo}
\eea
where 
\bea
f(\theta_2) &=& \frac{1}{(1+ c^2\alpha^2 \tan^2 \delta \sin^2 \theta_2)^2 }~.
\eea
The volume of the two-sphere may be computed analytically and is given by
\bea
\mathrm{vol} (S^2_\delta) &=& (4\pi \alpha^2 )\frac{\mathrm{arctanh} \left[ \frac{c\alpha \tan\delta}{\sqrt{1 + c^2\alpha^2 \tan^2 \delta }}\right]}{c\alpha \tan\delta \sqrt{1 + c^2\alpha^2 \tan^2 \delta }}~.
\eea
This is the volume of the two-sphere of the resolved conifold metric, times a function that is monotonically decreasing between one at $\delta=0$ and zero at $\delta = \pi/2$.
The non-K\"ahler deformation therefore squashes the two-sphere, and in the limit $\delta = \pi/2$ the volume goes to zero. 

This is an appropriate point to comment in more detail on the limit $\delta\rightarrow\pi/2$. In section \ref{global}, and then throughout the paper, 
we have chosen to identify coordinates in the rotated solution in such a way that for $\delta\in[0,\pi/2)$ the underlying manifold remains 
fixed. Indeed, we showed moreover in section \ref{complex} that the complex structure is then also invariant. 
Generically, this will be the only way to obtain a \emph{regular} supergravity solution for general $\delta$.\footnote{A general analysis 
here splits into different cases. However, the main issue is the loci $S$ where $\partial_\tau$ has fixed points. 
If one tries to take $w$ to have any period other than that of $\tau$, then one will obtain 
conical deficit singularities along $S$, leading to a singular supergravity solution.} However, with 
these global identifications the limit $\delta\rightarrow \pi/2$ is \emph{singular}, and not in fact the T-dual solution. 
The reason for this is simple: in general the T-dual solution requires a \emph{different} identification of 
coordinates, in order to obtain a regular supergravity solution. This is most clear when the T-dual solution 
has different topology; manifestly, one cannot then obtain the T-dual solution as a smooth limit  
$\delta\rightarrow\pi/2$ of smooth supergravity solutions, as the underlying manifolds have different topology. 
More computationally, the metric in the $w$ coordinates (\ref{wcoords}) is clearly degenerate 
at $\delta=\pi/2$. Instead the regular T-dual metric in this limit is described by giving a finite periodicity to the
original $\tau' = w\cos^2\delta$ coordinate.

Returning to the conifold example, 
the most interesting case is perhaps that with $a=c=d=1$, $b=0$. In this case, and as is well-known, the T-dual 
solution describes the back-reaction of two five-branes in flat spacetime. In the above coordinates, the 
two five-branes arise from the codimension four fixed point sets at $\{\theta_1=0,\theta_2=\pi\}$, 
$\{\theta_1=\pi, \theta_2=0\}$, which are two copies of $\R^2$ parametrized by $r, \psi$. 
These become the five-brane worldvolumes in the T-dual solution, while the minimal $S^2$ maps 
to an arc that joins the five-branes.

Let us briefly comment also on the $B$-field and gauge field. Notice that the pull-back of the $B$-field to the two-sphere vanishes. 
On the other hand, the pull-back of the gauge field is non-zero and reads
\bea
{\cal F}'|_{S^2} &=& c\alpha^2 \tan \delta \, \dd \left[  \frac{\sin^2 \theta_2 }{1+ c^2 \alpha^2 \tan^2 \delta \sin^2 \theta_2 } \dd \phi_2 \right]  ~.
\eea
However, the integral 
\bea
\int_{S^2} {\cal F}' = 0 ~.
\eea
Notice that if this integral were non-zero then the parameter $\delta$ would have been quantized. That it is not is in agreement with our general discussion 
of global properties of the deformed Calabi-Yau metrics in section \ref{global}. 

\section{Rotating heterotic string solutions: including $\alpha'$}
\label{alphaprime}

In this section we restore the $\alpha'$ of section \ref{heteroticsection}, 
and carefully describe in what 
sense the rotated solutions in the previous two sections are 
solutions to the low-energy limit of heterotic string theory.

\subsection{$\alpha'$ expansions}
\label{alpexp}

The supergravity fields in section \ref{heteroticsection} of course
 have an expansion in $\alpha'$. These are Taylor expansions 
around $\alpha'=0$, and in general we denote the coefficient of $\alpha'^n$ with a subscript $n$. 
Thus, for example, the metric is $g=g_0+\alpha'\, g_1+O(\alpha'^2)$. In particular, we note 
that the Bianchi identity (\ref{hetstringBianchi}) implies that
\bea
\diff H_0&=&0~.
\eea
A computation from the definitions shows that
\bea\label{identity}
{R}^+_{\ ijkl} - {R}^-_{\ klij} &=& \tfrac{1}{2}(\diff H)_{ijkl}~.
\eea
Notice the index structure. Thus to \emph{zeroth order} 
${R}^+_{0\, ijkl}= {R}^-_{0\, klij}$. The integrability condition 
for the gravitino equation immediately implies that
\bea
{R}^+_{0\, ijkl}\Gamma^{kl}\epsilon &=&0~,
\eea
which is the familiar statement of holonomy reduction for the Bismut connection $\nabla^+$, and thus from (\ref{identity}) we have that
\bea
{R}^-_{0\, klij}\Gamma^{kl}\epsilon&=&0~.
\eea
In other words, viewing $\mathscr{R}^-$ as the curvature of a connection on the tangent bundle, 
then this connection formally satisfies the last \emph{instanton equation} in (\ref{hetstringSUSY})
to zeroth order, with $\mathscr{F}$ replaced by $\mathscr{R}^-_0$. 

Also notice that in the Bianchi identity (\ref{hetstringBianchi}), and similarly the equations of motion 
(\ref{hetstringEOM}), one can effectively replace $\mathscr{R}^-$ by $\mathscr{R}^-_0$. 
This is simply because the $\alpha'$ corrections to the latter are at order $\alpha'^2$ 
in both (\ref{hetstringEOM}), (\ref{hetstringBianchi}),  and may hence be absorbed into 
$O(\alpha'^2)$. This has the important consequence that the connection that 
appears in these higher derivative terms satisfies the instanton equation. 
In appendix \ref{integrability} we show directly that if the supersymmetry 
equations and Bianchi identity are satisfied, up to and including 
first order in $\alpha'$ and with Bianchi identity 
\bea
\diff H &=& 2{\alpha'}\left(\mathrm{tr}\, \mathscr{F}\wedge\mathscr{F}-\mathrm{tr}\, \mathscr{R}^-_0\wedge \mathscr{R}^-_0\right)+O(\alpha'^2)~,
\eea
then the equations of motion (\ref{hetstringEOM}) follow.
For earlier results, 
see \cite{Gillard:2003jh,Ivanov:2009rh}.

The above comments lead to the following simple consequence. Suppose one has an exact supersymmetric solution 
to the above equations with $\mathscr{A}=0$ \emph{and} $\alpha'=0$. In other words, 
we have a supersymmetric type II solution, with 
 $\diff H=0$. We may now consider  turning on a non-Abelian gauge field by setting 
$\mathscr{F}=\mathscr{R}^-\equiv \mathscr{R}^-_0$ (at this point we have set $\alpha'=0$), which by virtue of the above discussion satisfies 
the instanton equation $\mathscr{F}_{ij}\Gamma^{ij}\epsilon=0$. 
Also, since $\mathscr{F}=\mathscr{R}^-$, the $O(\alpha')$ part of the Bianchi identity 
(\ref{hetstringBianchi}) is identitically zero. Reinstating $\alpha'\neq0$, we 
thus automatically obtain 
a supersymmetric solution to the heterotic string up to and including $O(\alpha')$. 
Moreover, this solution solves \emph{exactly} the above heterotic 
equations, with all $O(\alpha'^2)$ terms set to zero.\footnote{Of course, these terms 
are not zero in heterotic string theory.} This is a slightly modified form of the usual 
standard embedding. Notice that it is crucial that it is Hull's connection that 
appears in the higher derivative Bianchi identity, rather than say the Bismut or
Levi-Civita connection. In this case the non-Abelian gauge field takes values 
in a $Spin(6)\cong SU(4)\subset E_8$ subgroup. 

We conclude this subsection with an aside comment, which we feel is nevertheless important to make.
In the literature connections other than $\mathscr{R}^-$ 
have been taken in the Bianchi identity (\ref{hetstringBianchi}). In particular, the \emph{Chern 
connection} has been used. This requires some further comment. Suppose that 
the spacetime takes the product form $\R^{1,3}\times X$. Then, as reviewed in 
section \ref{nonka}, $X$ is a complex manifold equipped with in general a 
non-K\"ahler structure. The supersymmetry equations imply in particular 
that $\diff H$ has Hodge type $(2,2)$ with respect to the integrable complex structure. 
Thus in order to solve the Bianchi identity (\ref{hetstringBianchi}), 
up to and including order $\alpha'$, it follows that the curvature 
tensor that appears in this Bianchi identity should also have 
Hodge type $(2,2)$. This is indeed true of the Chern connection. However, 
the Chern connection is \emph{not} in general an instanton. 
In fact in the references in \cite{chern} the following identity is proven:
\bea\label{identityChern}
-\tfrac{1}{2}R^C_{\ ijkl}\omega^{ij}\omega^{kl}&=& -\tfrac{1}{2}R^+_{\ ijkl}
\omega^{ij}\omega^{kl} + C_{ijk}C^{ijk} + \tfrac{1}{4}(\diff H)_{ijkl}\omega^{ij}\omega^{kl}~.
\eea
Here $\omega$ is the type $(1,1)$ (non)-K\"ahler form, $R^C$ denotes the 
curvature of the Chern connection, and $C$ denotes its torsion. 
If the gravitino equation and Bianchi identity hold, up to order $\alpha'$, 
\emph{and} the Chern connection is an instanton (at zeroth order), then the zeroth order 
part of (\ref{identityChern}) implies that $C_0=0$. But
\bea
C_{ijk} &=& \tfrac{1}{2}\left(J^m_{\ \, i}(\diff\omega)_{mjk} + J^{m}_{\ \, j}(\diff\omega)_{imk}\right)~,
\eea
where $J$ denotes the complex structure tensor. Thus if $C_0=0$ then at zeroth order the solution is necessarily K\"ahler. 
In general this is not true! The problem with choosing the Chern connection 
in the Bianchi identity is then that supersymmetry and the Bianchi identity do \emph{not} 
imply the correct equations of motion -- again, we refer also to appendix \ref{integrability}.

\subsection{Rotations to first order in $\alpha'$}

Comparing the heterotic string equations of motion (\ref{hetstringEOM}) with 
the Abelian heterotic supergravity equations (\ref{hetEOM}), one sees that 
they are essentially the same, up to some small, but important, 
differences. Firstly, the heterotic string gauge field $\mathscr{A}$ 
is non-Abelian, while in the Abelian heterotic supergravity theory 
that embeds into type II, the gauge field $\mathcal{A}$ 
is Abelian. Of course, these are easily related by simply restricting $\mathscr{A}$
to lie in a $U(1)\cong SO(2)$ subgroup of the heterotic gauge group. 
Secondly, the gauge field in the heterotic string is accompanied by 
factors of $\alpha'$ in the action. Thirdly, the heterotic string 
also has higher derivative terms at this order in $\alpha'$. 

Let us now turn to our rotated Abelian heterotic solutions, focusing on the simpler case (\ref{rotatedCY})
in section \ref{CYcase}. In order not to cause confusion in what follows, we keep the 
primes in (\ref{rotatedCY}), now regarding these primes as denoting an Abelian 
heterotic supergravity solution, without any $\alpha'$.

We note first that when $\alpha'=0$, the gauge field 
no longer appears in the low-energy action of the heterotic string. 
This suggests that in order to introduce an $\alpha'$ dependence 
into the rotated solutions (\ref{rotatedCY}), we set the rotation angle
\bea\label{deltaalpha}
\delta &=& \sqrt{\alpha'}\, \lambda~.
\eea
This is simply because when $\alpha'=0$ the solution should have zero gauge field, which implies 
that $\delta$ should also be zero. We then define
\bea\label{As}
\mathscr{A} &=& \frac{1}{\sqrt{\alpha'}}\mathcal{A}'~,\nn\\
&=& -\lambda\left[\diff\tau'-\frac{1}{V}(\diff\tau'+A)\right]+O(\alpha')~.
\eea
Here one should be more precise about how the $U(1)\cong SO(2)$ gauge field $\mathcal{A}'$ is embedded into the 
heterotic gauge group. Obviously, there are choices. For example, we may more precisely write
\bea\label{inbed}
\mathscr{A} &=& -\lambda\left[\diff\tau'-\frac{1}{V}(\diff\tau'+A)\right]\left(\begin{array}{cccc}0& -\tfrac{\ii}{\sqrt{2}} & & \\ \tfrac{\ii}{\sqrt{2}} & 0 & & \\ & & \ddots & \\ &&& 0\end{array}\right)+O(\alpha')~.
\eea
The purely imaginary, skew-symmetric matrix here embeds the $U(1)$ gauge field $\mathcal{A}'$ as 
an $SO(2)$ subgroup of either $SO(32)$ or $SO(16)\subset E_8$. However, most of what we say will be independent 
of these details, and we thus suppress this for the time being.
The explicit introduction of 
 $\sqrt{\alpha'}$ in  (\ref{As}) and normalization in (\ref{inbed}) ensures 
that the kinetic terms are then related as $\mathcal{F}'_{ij}\mathcal{F}'^{ij} = 
\alpha'\, \mathrm{tr}\, \mathscr{F}_{ij}\mathscr{F}^{ij}$, and similarly now the 
Bianchi identity $\diff H'=\mathcal{F}'\wedge\mathcal{F}'$ 
becomes $\diff H'=\alpha'\, \mathrm{tr}\, \mathscr{F}\wedge\mathscr{F}$. 
Since the other fields already appear at order zero, we simply set 
$g=g'$, $H=H'$ and $\Phi=\Phi'$, where the primed fields now 
depend on $\alpha'$ via (\ref{deltaalpha}). The full solution is then easily computed to be
\bea\label{alphasoln}
g&=& g_{\mathrm{CY}} - \frac{2\alpha'\lambda^2}{V}\left(\diff\tau'+A\right)\left(\diff\tau'+A-\tfrac{V}{2}\diff\tau'\right) + O(\alpha'^2)~,\nn\\
\me^{2\Phi} &=& \me^{2\Phi_0}\left[1-\frac{\alpha'\lambda^2}{V}\left(1-\tfrac{V}{2}\right)\right]+O(\alpha'^2)~,\nn\\
\mathscr{A}&=& -\lambda\left[\diff\tau'-\frac{1}{V}(\diff\tau'+A)\right]+O(\alpha')~,\nn\\
B&=& -\frac{\alpha'\lambda^2}{V}\diff\tau'\wedge A+O(\alpha'^2)~.
\eea
This is a one-parameter family of deformations of the original Calabi-Yau metric, with deformation parameter $\lambda$.
At zeroth order the solution is Calabi-Yau, with $B_0=0$. Notice that the zeroth order gauge field $\mathscr{A}_0$
is non-zero; however, at this order the gauge field does not appear in the action. By construction, 
(\ref{alphasoln}), \emph{including} the infinite $\alpha'$ expansion coming from the original solution (\ref{rotatedCY}), solves the supersymmetry equations (\ref{hetstringSUSY}) \emph{exactly} up to and including first order in $\alpha'$. However, from a physical perspective one should truncate the solution at order $\alpha'^2$, as we have done in (\ref{alphasoln}), since 
the supersymmetry equations will receive corrections at this order. 

However, the Bianchi identity (\ref{hetstringBianchi}) is solved
\emph{without} the higher derivative terms in $\mathscr{R}^-$. We may try to incorporate this 
using a modification of the standard embedding. Notice that $\mathscr{R}^-_0=\mathscr{R}_{\mathrm{CY}}$ 
is simply the curvature tensor of the original Calabi-Yau metric, and as such indeed solves the instanton 
equation at \emph{zeroth order}. This is also an $SU(3)$ connection. Perhaps the minimal choice 
of ``modified standard embedding'' is then to write the following $SU(4)\subset E_8$ gauge field 
\bea\label{fullgauge}
\mathscr{A}_{\mathrm{MSE}} &=& \frac{1}{\sqrt{\alpha'}}\mathcal{A}' \, {\cal Q}_4+ \omega^{\mathrm{spin}}_{\mathrm{CY}}~,
\eea
where 
\bea
{\cal Q}_4 & = & \left(\begin{array}{cccccc}\tfrac{\sqrt{3}}{2} & 0&0&0&&\\ 0&-\tfrac{1}{2\sqrt{3}}&0&0&&\\ 0&0&-\tfrac{1}{2\sqrt{3}}&0&&\\ 0&0&0&-\tfrac{1}{2\sqrt{3}}\\ &&&&\ddots &\\ &&&&&0\end{array}\right)~.
\eea
Here the spin connection of the Calabi-Yau manifold
$\omega^{\mathrm{spin}}_{\mathrm{CY}}$, which is an $SU(3)$ connection, is understood to be embedded in 
the $3\times 3$ block that commutes with ${\cal Q}_4$.
Then we have that 
\bea
\mathscr{F}_{\mathrm{MSE}} & = & \mathscr{R}_\mathrm{CY} + \frac{1}{\sqrt{\alpha'}}{\cal Q}_4\, \mathcal{F}'~,
\eea
and the normalizations guarantee that
\bea\label{fullBianchi}
\diff H &=& \mathcal{F}'\wedge\mathcal{F}' \nn \\ & = &  
\alpha'\left(\mathrm{tr}\, \mathscr{F}_{\mathrm{MSE}}\wedge\mathscr{F}_{\mathrm{MSE}} - 
\mathrm{tr}\, \mathscr{R}^-\wedge\mathscr{R}^-\right) + O(\alpha'^2)~,
 \eea
 and thus we solve the Bianchi identity at the appropriate order. Notice here that it is crucial that 
 the $U(1)$ gauge field commutes with the $SU(3)$ spin connection of the original Calabi-Yau metric in (\ref{fullgauge}), 
 ensuring that these terms do not mix in the wedge product and associated matrix multiplication. Of course, 
 it is also important that both are traceless.
 
 The first term in (\ref{fullgauge}) solves the instanton equation up to and including $O(\alpha')$, by 
 construction. However, $\mathscr{R}_{\mathrm{CY}}$ is only an instanton \emph{at zeroth order}. 
Thus at this point we have solved all equations up to and including $O(\alpha')$, 
\emph{except} the instanton equation, which is solved only at zeroth order. 
However, notice that the gauge field $\mathscr{A}$ enters the action (\ref{hetstringaction}) 
already at order $\alpha'$. Thus an $O(\alpha')$ correction to $\mathscr{A}$ enters 
the action at \emph{second order}, while the corresponing $O(\alpha')$ corrections 
to the metric, dilaton and $B$-field enter at first order. In this sense, we 
already have a solution up to and including $O(\alpha')$. Nevertheless, we 
shall briefly comment further on higher order corrections below.

The gauge field (\ref{fullgauge}) lies in a $U(1)\times SU(3)\subset SU(4)\subset E_8$ 
subgroup of the full heterotic gauge group. The maximal commuting subgroup in $E_8$ is 
then $SO(10)\times U(1)$, where $SO(10)$ is the commutant of $SU(4)$ in $E_8$ and the $U(1)$ factor is the same as that in (\ref{fullgauge}).
Indeed, an alternative way to think about this is that the initial solution with $\delta=0$ has only an $SU(3)$ gauge field turned on, via the standard
embedding, and as such has commutant $E_6$. Then the rotation turns on an additional $U(1)$ gauge field which breaks this to 
$SO(10)\times U(1)$.
 Although our solutions here 
are non-compact, one might imagine that these are good \emph{local models} 
for a compact non-K\"ahler heterotic solution. For example, the conifold 
metric is believed to model a neighbourhood of a compact Calabi-Yau manifold 
near to a conifold transition, and our rotation then applies directly 
to the resolved and deformed conifolds. In this setting, and assuming our gauge field 
extends globally as a $U(1)\times SU(3)$ or $SU(4)$ gauge field, then 
the low-energy gauge group in $\R^{1,3}$ is $SO(10)$. In particular, 
in the former case the $U(1)$ commutant obtains a mass 
via the Green-Schwarz mechanism -- at least, this is so in the Calabi-Yau case, as discussed, for example, in \cite{Blumenhagen:2006ux}. 
Thus the rotation effectively breaks the initial $E_6$ gauge group to $SO(10)$.

A slight generalization of the above contruction can be given when the initial solution has non-zero $H_0$. 
As explained in section
\ref{alpexp}, via the standard embedding using Hull's connection in this case the initial non-Abelian gauge field lies in an $SU(4)$ subgroup 
(the holonomy of the $\omega^-_0$ connection) of $E_8$. 
The rotation turns on an Abelian gauge field via the embedding 
\bea
\mathscr{A}_{\mathrm{MSE}} & = & \frac{1}{\sqrt{\alpha'}} \mathcal{A}' \, {\cal Q}_5 + \omega_0^-~,
\eea
where ${\cal Q}_5 = \tfrac{1}{\sqrt{20}} \mathrm{diag}(4,-1,-1,-1,-1,0,\dots)$, thus breaking the low-energy gauge group to  $SU(5)$.
In addition to \cite{Maldacena:2000yy}, solutions amenable to this construction include the $T^2$ bundles over conformally hyper-K\"ahler 
two-folds, presented explicitly in \cite{Goldstein:2002pg,Gauntlett:2003cy}. 

We conclude this section with a brief comment on the possibility of
correcting (\ref{fullgauge}) at order $\alpha'$ to solve the instanton equation at this order. Notice that 
this correction, whatever it is, does \emph{not} affect the Bianchi identity (\ref{fullBianchi}), 
since such a correction enters at $O(\alpha'^2)$ in the Bianchi identity. 
The philosophy from here  follows the important, but 
relatively unknown, paper by Witten and Witten \cite{Wittens}.\footnote{Reference \cite{Wittens} is crucial 
for the consistency of Calabi-Yau compactifications in which one takes a Hermitian-Yang-Mills 
gauge field that is \emph{not} the standard embedding. Such a solution leads to 
$\diff H_1\neq 0$, which means that the metric is Calabi-Yau only to leading order. Despite 
the large number of papers on this subject, \cite{Wittens} only has 21 references on SPIRES at the time of writing.} 

We write the non-Abelian gauge field, up to and including order $\alpha'$, as
\bea\label{fullfull}
\mathscr{A}^{(1)} &=& \mathscr{A}_{\mathrm{MSE}} + \alpha' S~,
\eea
where $S$ is an $\alpha'$-independent one-form with values in the Lie algebra of $SU(4)$. Notice here 
that $\mathscr{A}_{\mathrm{MSE}}$, as we have defined it, contains all powers of $\alpha'$. 
In particular, the (perhaps confusing) notation above implies that 
$\mathscr{A}^{(1)}_0 = \mathscr{A}_{\mathrm{MSE}\, 0}$ and
$\mathscr{A}^{(1)}_1=\mathscr{A}_{\mathrm{MSE}\, 1}+S$
Expanding everything in powers of $\alpha'$
and collecting all linear terms we obtain the equations
\bea
\Omega_0 \wedge  \mathscr{F}^{(1)}_{1} & =  & - \Omega_1 \wedge  \mathscr{F}^{(1)}_{0}~,\nonumber\\
\omega_0 \wedge \omega_0 \wedge   \mathscr{F}^{(1)}_{1} & = &  -2 \omega_0 \wedge \omega_1 \wedge  \mathscr{F}^{(1)}_{0}~.
\label{qww}
\eea
The left hand sides of these equations are just the usual instanton equations, and vanish if one replaces $\mathscr{F}^{(1)}_{1}$
by $\mathscr{F}^{(1)}_{0}$, by construction. The right hand sides are then 
``source'' terms. One could decompose $\mathscr{F}^{(1)}_{1}$ into irreducible representations with respect to the zeroth order structure, and use the equations above to determine its components. However, $\mathscr{F}^{(1)}_{1}$ is not arbitrary, since the Bianchi identity for the gauge field implies 
\bea
\mathscr{F}^{(1)}_{1} & = & \diff \mathscr{A}^{(1)}_{1} + \mathscr{A}_{\mathrm{MSE}\, 0}\wedge \mathscr{A}^{(1)}_{1} + \mathscr{A}^{(1)}_{1}\wedge\mathscr{A}_{\mathrm{MSE}\, 0}~.
\eea
Notice that the equations (\ref{qww}) are linear, first order PDEs for the connection $S$,
where all other terms are known from the rotated 
solution. It would be interesting to investigate this system further.

\section{Discussion}
\label{theend}

In this paper we have studied a solution-generating transformation in the context of heterotic supergravity, 
with a non-trivial $U(1)$ 
gauge field in the Cartan subgroup of $E_8 \times E_8$ or $SO(32)$. In particular, we have discussed 
how this transformation preserves supersymmetry. In the case of 
backgrounds of the form $\R^{1,3}\times X$ we have explicitly shown that 
solutions obtained as rotations of Calabi-Yau geometries satisfy the non-K\"ahler equations, and 
that the complex structure remains invariant.

One of the main observations we have made is that heterotic solutions with an Abelian gauge field can be formally mapped to type II solutions, 
where the heterotic gauge field becomes a component of the metric in an internal space of one dimension higher (see also \cite{Gauntlett:2003cy}). 
Based on this, we have seen  that an $O(2,1)$ subgroup of the (Abelian) heterotic duality group $O(d+16,d)$ is effectively embedded into an $O(2,2)$ subgroup 
of the type II T-duality group. We thus realize the heterotic transformation  as a simple 
combination of rotations and ordinary Buscher T-duality in type II theories. This relationship perhaps deserves to be further studied. 
For example, it would be interesting to investigate whether a suitable notion of heterotic generalized geometry exists based on the 
$O(d+16,d)$ duality group, generalizing the generalized geometry that puts the metric and $B$-field on the same footing in type II.

In the context of heterotic string theory, taking into proper account the parameter $\alpha'$ leads to some modifications of the discussion. 
Firstly, although the solution-generating formulae formally contain an infinite series\footnote{In \cite{Hassan:1991mq} it is suggested that 
the symmetry should be valid at all orders in $\alpha'$, although the precise transformation of the fields may be corrected.} 
in powers of $\alpha'$, the equations of motion, supersymmetry variations and Bianchi identities for the heterotic theory 
are only known up to some low order in the $\alpha'$ expansion \cite{roo}. Therefore, the transformation only makes sense if the expansions 
are truncated at the appropriate order. In addition, the duality transformations do not apply to 
non-Abelian gauge fields. As we discussed, it turns out that it is quite simple to remedy this, and obtain supersymmetric 
heterotic solutions at \emph{first order} in $\alpha'$, including a non-Abelian gauge field with a slightly 
modified standard embedding. Thus, given any (non-compact) Calabi-Yau geometry, we have constructed non-K\"ahler solutions, breaking the heterotic gauge group 
from $E_6$ (for the Calabi-Yau with standard embedding) to the GUT gauge group  $SO(10)$.  

Going back to the type II setting, the transformation we discussed is then \emph{exact}, 
since the gauge field is just a metric component and hence appears at the same (lowest) order in $\alpha'$ here. 
We may then start in type II with the direct product of a Calabi-Yau geometry with a circle, and the transformation will ``twist'' 
this circle over the base Calabi-Yau (although topologically this twisting is always trivial), with the base becoming precisely 
a non-K\"ahler geometry. Furthermore, we can start with any supersymmetric solution with non-trivial $B$-field, and 
also perform the transformation.\footnote{In this case the twisting could change the global topology of the transformed solution.} 
For example, one could apply this rotation to the Maldacena-Nu\~nez solution \cite{Maldacena:2000yy}.

Notice that if we start with a type IIB  solution of the form
$\R^{1,2}\times S^1_z \times X$, a T-duality  along  $S^1_z$ will give  a $\R^{1,2}\times \tilde S^1_z \times X$ geometry in type IIA. 
We can then peform a rotation in the $(z,\tau)$ plane, and finally another T-duality along the rotated $\tilde S^1_z$.
Let us call this a TrT transformation. If we peform a further rotation of the final type IIB solution, we have simply composed
a T-duality with our solution-generating transformation. However, from a \emph{physical} point of view, the second rotation does nothing, and hence 
effectively we have performed a TrT transformation. This is (locally) equivalent to a TsT transformation, as shown in appendix \ref{TrTappendix}. 
The latter can be interpreted in the dual field theory (when this exists) as a \emph{dipole deformation} \cite{Lunin:2005jy, Gursoy:2005cn}.
In these cases one could reinterpret our results in the dual field theory. 
It would be interesting to explore applications of our results to five-brane solutions representing gravity duals of field theories.  

\subsection*{Acknowledgements}
We thank Rhys Davies, Chris Hull, Carlos Nu\~nez and George Papadopoulos for discussions and useful comments. 
We are particularly grateful to George Papadopoulos for comments on a draft version of this paper. 
D. M. is partially supported by an EPSRC Advanced Fellowship EP/D07150X/3.
J. F. S. is  supported by a Royal Society University Research Fellowship.

%\appendix 

\begin{appendix}

\section{An integrability result}
\label{integrability}

In this appendix we prove an integrability result in heterotic supergravity, 
extending the result in the appendix of  \cite{Gauntlett:2002sc} to include  the Bianchi identity corrected at first order in $\alpha'$. 
This is a simplified version of the proof given in \cite{Ivanov:2009rh}. 
The details of the proofs of the latter reference differ depending on the dimension of the internal manifold, namely on the 
form of the ten-dimensional spacetime, which is taken to be of the product form $\R^{1,p}\times X_{9-p}$. 
In contrast, our ``spinorial'' proof is valid in any dimension $p$. 

Let us record here the equations of motion at first order in $\alpha'$
\bea\label{hetstringEOMagain}
{R}_{ij}+2\nabla_i\nabla_j\Phi-\tfrac{1}{4}{H}_{ikl}{H}_{j}^{\ kl} - 2{\alpha'}\, \mathrm{tr}\, \mathscr{F}_{ik}\mathscr{F}_{j}^{\ k}
 \nn \\+2 \alpha'\, \mathrm{tr}\,  \mathscr{R}_{ ik}\mathscr{R}_j{}^{k}&=&0~,\\
\nabla^2(\me^{-2\Phi})-\tfrac{1}{6}\me^{-2\Phi}\, {H}_{ijk}{H}^{ijk} -{\alpha'}\, \me^{-2\Phi} \mathrm{tr}\,  \mathscr{F}_{ij}\mathscr{F}^{ij} 
 \nn \\+ {\alpha'}\, \me^{-2\Phi} \mathrm{tr}\,  \mathscr{R}_{ ij}\mathscr{R}^{ ij}&=&0~,\label{dalek}\\
\nabla^i\left(\me^{-2\Phi}\, {H}_{ijk}\right)&=&0~,\\
\nabla^{+\, i}\left(\me^{-2\Phi}\mathscr{F}_{ij}\right) &=&0~,
\eea
and the modified Bianchi identity 
\bea\label{Bianchihigh}
\diff H = 2 \alpha' \left(\mathrm{tr}\, \mathscr{F}\wedge \mathscr{F} - \mathrm{tr}\,  \mathscr{R}  \wedge  \mathscr{R} \right) ~.
\eea
In this appendix we do not specify the connection $\omega$ used to compute the curvature 
$\mathscr{R}= \dd \omega + \omega \wedge \omega$. We will see that the integrability results are compatible with   
$\mathscr{R}=\mathscr{R}^-_0$, namely the zeroth order part of the Hull curvature. 

We shall not explicitly write the gauge group indices on the field 
strength $\mathscr{F}_{ij}{}^{ab}$. Similarly, we denote by $\mathscr{R}_{ij}$ the curvature two-form 
\bea
\mathscr{ R}_{ij}{}^{ab} &=& R_{ijkl}e^{a k}e^{b l}~,
\eea
where $R_{ijkl}$ is the Riemann tensor computed with the given connection. 
We record also the supersymmetry equations in these conventions:  
\bea
\left(\nabla_i  + \tfrac{1}{8}H_{ijk}\Gamma^{jk} \right) \epsilon & = & 0~,\label{gravi}\\
\left( \Gamma^i\de_i \Phi + \tfrac{1}{12} H_{ijk}\Gamma^{ijk} \right) \epsilon & = & 0~, \label{dilat}\\
\mathscr{F}_{ij}\Gamma^{ij}  \epsilon & = & 0~.
\label{gaugi}
\eea
Notice we are using exactly the same conventions as \cite{Gauntlett:2002sc}. We can then use  equation (8.7) 
of this reference, which we record here:
\bea
\left(\nabla^2 \Phi - 2 (\nabla \Phi)^2 + \tfrac{1}{12} H_{ijk}H^{ijk} +    \tfrac{\alpha'}{2}\mathrm{tr}\, \mathscr{F}_{ij} \mathscr{F}^{ij}  \right) \epsilon \ =\ \nn\\[2mm]
~~~{ ~~ }~~~ - \tfrac{1}{48}\left(\dd H - 2 \alpha' \mathrm{tr}\, \mathscr{F} \wedge  \mathscr{F} \right)_{ijkl} \Gamma^{ijkl}\epsilon  
- \tfrac{1}{4}\me^{2\Phi} \nabla^i \left(\me^{-2\Phi}H_{ijk}\right)  \Gamma^{jk}\epsilon~.
\label{liftedid}
\eea
This is derived using only the dilatino (\ref{dilat}) and gaugino (\ref{gaugi}) supersymmetry equations. 
Let us now assume that the spacetime is of the form $\R^{1,p}\times X_{9-p}$. 
Then one of the equations following from supersymmetry 
is the calibration condition \cite{Gauntlett:2003cy}
\bea\label{calibrate}
 \me^{-2\Phi} *_{9-p} H & = &-  \diff (\me^{-2\Phi} \Xi )~.
\eea
Here $\Xi$ is a $G$-invariant form specifying, at least in part, the related $G$-structure. 
For example, when $p=3$ then $\Xi=\omega$ is the type $(1,1)$ form for the associated $G=SU(3)$ 
structure, while when $p=2$ then $\Xi=\phi$ is the associative three-form for the associated $G=G_2$ structure; 
a complete discussion may be found in \cite{Gauntlett:2003cy}.
Equation (\ref{calibrate}) then automatically implies the $H$ equation of motion. Hence the last term in (\ref{liftedid}) is zero.
Now, if we assume that the Bianchi identity (\ref{Bianchihigh}) and the dilaton equation of motion (\ref{dalek}) hold,  
substituting them into (\ref{liftedid}) we obtain 
\bea
  \tfrac{\alpha'}{2}\mathrm{tr}\, \mathscr{R}_{ij} \mathscr{R}^{ij}   \epsilon &=&
\tfrac{\alpha'}{24} \left( \mathrm{tr}\, \mathscr{R} \wedge  \mathscr{R} \right)_{ijkl} \Gamma^{ijkl}\epsilon  ~.
\eea
Using the identity
\bea
\{\Gamma_{ij}, \Gamma^{kl} \}& =&  2 \Gamma_{ij}{}^{kl} - 4 \delta_{ij}^{kl}~,
\eea
the latter equation can be written as
\bea
\{\Gamma^{ij}, \Gamma^{kl} \} \mathrm{tr}\, (\mathscr{R}_{ij} \mathscr{R}_{kl} ) \epsilon & =& 0~.\eea
Multiplying on the left by $\epsilon^\dagger$, defining $P = \Gamma^{ij} \mathscr{R}_{ij}$ and noting\footnote{This is true if the $\Gamma^i$ are Hermitian, or anti-Hermitian, 
which can always be arranged in the cases of interest.} that $P^\dagger = - P$, we obtain
\bea
\mathrm{tr}\, [ (P\epsilon)^\dagger P\epsilon]&=&0~.
\eea
However, since the trace is positive definite this implies that $P\epsilon =0$, which is the instanton equation
\bea
 \mathscr{R}_{ij}\Gamma^{ij}  \epsilon &=& 0 ~.
\eea
We conclude, using also the remaining results in the appendix of reference \cite{Gauntlett:2002sc}, that the supersymmetry variations (\ref{gravi}), (\ref{dilat}), (\ref{gaugi}), together with 
the Bianchi identity (\ref{Bianchihigh}), imply the equations of motion (\ref{hetstringEOMagain})
if and only if the curvature of the connection used in (\ref{Bianchihigh}) and (\ref{hetstringEOMagain}) is an \emph{instanton}.
In particular, to this order in the $\alpha'$ expansion, Hull's curvature  $\mathscr{R}^-$ is singled out by compatibility of supersymmetry and 
the equations of motion. 

\section{TrT = TsT}
\label{TrTappendix}

In this appendix we show that, at least locally,  a TrT transformation is equivalent to a TsT transformation, 
for any configuration without RR fields. Presumably  this result extends to configurations with RR fields, 
although we have not examined this.
The following computation is a straightforward, but slightly tedious, application of the T-duality rules. 

We begin  with a solution
\be
\begin{aligned}
\diff s^2 \ =\  & \,  f_1 (\diff x_1 + a\diff x_2 + A_1)^2 +  f_2 (\diff x_2  + A_2)^2 + \diff s^2_\perp~,\\
B \ =\  & \, B_1 \wedge \diff x_1 + B_2 \wedge \diff x_2  + b  \, \diff x_1 \wedge \diff x_2 + B_\perp~,
\end{aligned}
\label{start}
\ee
where the metric is in string frame. We also have a non-trivial dilaton field $\Phi$. $\de/ \de x_1 $ and $\de /\de x_2$ are Killing vectors. 
The metric is written in a form adapted to performing a T-duality along the $\de/ \de x_1 $ direction. We then have that  $A_1$, $A_2$ and
$B_1$, $B_2$ have zero contractions with $\de/\de x_1$ and $\de/\de x_2$, and moreover these one-forms, together with the 
 functions $f_1, f_2, a,b$, are independent of $x_1, x_2$.
We compute in turn the TrT and TsT transformations and compare the results at the end. 

For the TrT transformation we start by performing a T-duality along
$\de/ \de x_1$, followed by a \emph{{\bf r}otation} of the (T-dualized) Killing coordinates $x_1, x_2$:
\be
\begin{aligned}
x_1&=&  \cos \theta\, \check x_1 - \sin\theta\, \check x_2~, \\
x_2 &=&  \sin\theta\, \check x_1 + \cos\theta\,  \check x_2~.
\end{aligned}
\ee
We then perform a T-duality along $\de / \de \check x_1$, denoting the T-dualized coordinates by $\hat x_1, \hat x_2$. The TrT transformed 
metric and dilaton then read
\be
\begin{aligned}
\diff  s'^2 \ =\  & \, f^{-1}\Big[  f_1 \left[\diff  x'_1 + a\diff x'_2  + (c - s b ) A_1    - s (a B_1 - B_2) \right]^2 \\
& ~~~ +f_2 \left[\diff x'_2  + (c - s b )A_2 - s B_1\right]^2 \Big] +  \diff s^2_\perp~,\\
\me^{2\Phi'} \ = \  & \frac{\me^{2 \Phi} }{f}~,~~~~~~~~~~~~~~~~~~~~~~f \ \equiv \ (c  -  s b )^2 + s^2 f_1 f_2~,
\end{aligned}
\ee
where we have used the shorthand notation $s=\sin\theta$, $c=\cos \theta$. The transformed $B$-field is
\be
\begin{aligned}
\hat B & \ =\  B_\perp - B_1 \wedge A_1 + \diff x'_2 \wedge \big[ (s+ c b) A_1 + c (aB_1 - B_2) \big]\\
& + f^{-1} \big[ (c-s b) [B_1 - (s+cb ) \diff x'_2 ]+ s f_1 f_2 (c \diff x'_2 + A_2) \big]\\
& ~~~~~~~~\wedge \big[ \diff x'_1 + a \diff  x'_2  + (c - s b ) A_1    - s (a B_1 - B_2) \big]~.
\end{aligned}
\ee

For the TsT transformation we start again from the solution (\ref{start}) and perform a T-duality along
$\de/ \de x_1$. We then \emph{{\bf s}hift} the (T-dualized) Killing 
coordinates $x_1, x_2$:
\be
\begin{aligned}
x_1 & \ =\   \check x_1~, \\
x_2 & \ =\  \check x_2 + \gamma \check x_1~.
\end{aligned}
\ee
Finally, we perform a T-duality along $\de / \de \check x_1$, denoting the T-dualized coordinates by $ x'_1, x'_2$. The TsT transformed 
metric and dilaton then read
\be
\begin{aligned}
\diff  s'^2 \ =\ & \, h^{-1}\Big[  f_1 \left[\diff x'_1 + a\diff x'_2  + (1 - \gamma b ) A_1    - \gamma (a B_1 - B_2) \right]^2 \\
& ~~~ +f_2 \left[\diff x'_2  + (1 - \gamma b )A_2 - \gamma B_1\right]^2 \Big] +  \diff s^2_\perp~,\\
\me^{2\Phi'}\  =\ & \frac{\me^{2 \Phi} }{h}~,~~~~~~~~~~~~~~~~~~~~~~h \ \equiv\  (1  -  \gamma b )^2 + \gamma^2 f_1 f_2~.
\end{aligned}
\ee
The transformed $B$-field is
\be
\begin{aligned}
\hat B & \ =\ B_\perp - B_1 \wedge A_1 + \diff x'_2 \wedge \big[  b A_1 + aB_1 - B_2 \big]\\
& + h^{-1} \big[ (1-\gamma b) [ B_1 - b\diff  x'_2] + \gamma f_1 f_2 (\diff x'_2 + A_2)\big] \\
& ~~~~~~~~\wedge \big[ \diff x'_1 + a\diff x'_2  + (1 - \gamma b ) A_1    - \gamma (a B_1 - B_2)  \big] ~.
\end{aligned}
\ee

Let us now compare the results of the two transformations. We see that the 
metrics agree precisely if we rescale the coordinates of the TrT solution 
as $x'_{i\, \mathrm{TrT}} = \cos \theta\,    x'_{i\, \mathrm{TsT}}$
and identify $\gamma = \tan \theta$. The dilatons are then related via 
\be
\me^{2 \Phi'}|_\mathrm{TsT}\ =\ \cos^2 \theta \, \me^{2\Phi'}|_\mathrm{TrT}  ~.
\ee
For the $B$-fields one immediately sees that some terms in the two expressions clearly match, but some others apparently differ.
A calculation  shows that the two expressions are indeed different, but related by 
\be
 B'_\mathrm{TsT} \ =\   B'_\mathrm{TrT} - \sin\theta \cos\theta  \, \diff  x'_1 \wedge \diff x'_2 ~.
\ee
The difference is a closed two-form, and hence the two configurations differ by only a flat $B$-field.

\end{appendix}

\bibliographystyle{my-h-elsevier}

\begin{thebibliography}{10}
\bibitem{reidfantasy}
M.~Reid,
``The moduli space of 3-folds with $K=0$ may nevertheless be irreducible,''
 Mathematische  Annalen, Volume 278,  Numbers 1-4, 329-334
\end{thebibliography}

\end{document}